\documentstyle[12pt,epsf,epsfig,rotating]{article}
\input axodraw.sty

\begin{document}
\title{Search for Standard Higgs Boson at Supercolliders}
\author{N.V.Krasnikov and V.A.Matveev \\
INR RAS, Moscow 117312}

\date{September, 1999}
\maketitle
\begin{abstract}
We review the standard Higgs boson physics and the search for 
standard Higgs boson at LEP and LHC supercolliders 

\end{abstract}

\newpage

{\bf Content}

1. {\bf Introduction}

2. {\bf The Lagrangian of the Standard Model}

3. {\bf Higgs boson decays}

4. {\bf Indirect bounds on the Higgs boson mass}

5. {\bf Search for Higgs boson at LEP}

6. {\bf Higgs boson production at hadron supercolliders}

7. {\bf  Search for Higgs boson at TEVATRON}

8. {\bf LHC: CMS and ATLAS detectors}

9. {\bf Search for Higgs boson at LHC}

10. {\bf Conclusion}

\newpage
\section{Introduction} 

\begin{flushright}
     \small{This paper is devoted to the memory of \\
     our teacher Nikolai Nikolaevich Bogolyubov \\
     whose 90-th year jubilee is celebrated by \\
     the physics and mathematical community}
\end{flushright}

The Standard Model which describes within an unprecendental scale of 
energies and distances the strong and electroweak interactions of elementary 
particles relays on a few basic principles - the renormalizability, 
the gauge invariance and the spontaneous  breaking of the underlying 
gauge symmetry. The principle of the renormalizability which is considered 
often as something lying beyond the limits of experimental test is in 
fact one of the most important (if not the major) ingredients of the 
quantum field theory.

The requirement of renormalizabilty which content and deep meaning were 
uncovered in the fundamental textbook by N.N.Bogolyubov and D.V.Shirkov 
\cite{1} plays the central role in the construction and classification 
of the field theoretic models. They split in general on two classes. 

In the renormalizable models the ultraviolet divergences of the    
radiative corrections are under mathematically rigorous control due to the 
famous Bogolyubov-Parasiuk theorem \cite{2}. 
These models which preserve their locality in all orders of the 
perturbation theory are characterised by a finite number of relevant 
dimensionless coupling constants whose dependence on an arbitrary 
dimensional normalization parameter is described by the renormalization 
group \cite{1}. These so-called the ``running'' coupling constants 
depending on the model may have or the asymptotic freedom behaviour at 
large momenta (as for non-abelian gauge theories) or like in 
quantum electrodynamics with an abelian gauge symmetry reveal the 
growth of the effective coupling constant  in the ultraviolet region. 

The second class of field theoretical models - the non-renormalizable 
models have a very serious drawback which makes them useless for 
description of particle interactions at the present level of knowledge. 
First of all, the non-renormalizable models have infinite number 
of divergent matrix elements that requires as a consequence an introduction of 
an infinite number of interaction vertecies and dimensional coupling 
constants. What is more important the nonrenormalizable theories are 
nonlocal and depend on the infinite number of unknown functions \cite{1,3}. 
This follows from the fact  that the vertecies of the nonrenormalizable 
models contain an arbitrary high derivatives of the field operators. 
Thus the predictive power of nonrenormalizable models is 
close to zero. An imaging world described by such a theory seems to be 
extremely complicated unlike what we learn from studying particle 
interactions and evolution of the Universe at least until the present. 
\footnote{However we should not ignore the fact that the distinction between 
renormalizable and nonrenormalizable theories is evident only within 
perturbation theory.}     
                                               
The Weinberg-Salam  model \cite{4} 
of the electroweak interactions belongs to the first 
class of the field theories. The major ingredient of this  model 
which experimental test is the target of the world-wide search 
programme is the presence of the scalar multiplet with 
nontrivial vacuum condensate (the Higgs boson \cite{5}). 
The nonzero vacuum condensate does not affect the small distance 
behaviour of particle interactions that  allows to solve the 
problem of mass generation for $W$- and $Z$- vector bosons without 
conflict with  the renormalizabilty of the theory. The spontaneous 
breaking of the gauge symmetry in the Weinberg-Salam model is 
 a consequence  of the degeneration of the ground state in 
the presence of the boson condensate  - in precise analogy with 
the theory of the superfluidity \cite{6}. 
   
In Weinberg-Salam model a complex isodoublet 
scalar field is introduced  through self-interactions; this acquires 
non-vanishing vacuum expectation value, breaking spontaneously the 
electroweak gauge group $SU(2)_L \otimes U(1)$ to the electromagnetic 
$U(1)_{EM}$ gauge group. The interactions of the gauge bosons and fermions 
with  the background field generate the masses of these particles. One 
component of the  scalar isodoublet Higgs field is not absorbed in the 
longitudinal components of the vector $W$ and $Z$ bosons, manifesting 
itself as the physical Higgs particle $h$ \footnote{Note that very 
often standard Higgs boson is denoted by capital letter H.}.
It should be stressed that the Higgs mechanism is the only way to construct 
the renormalizable theory of the electroweak interactions. Therefore the 
 discovery of  the single missing ingredient of the Weinberg-Salam model 
 - the Higgs boson will be in some sense the experimental ``proof'' of the 
renormalizability of the electroweak interactions.  
There are no doubts that at present 
the main supergoal in high energy experimental physics is the search for 
the Higgs boson. 

In this paper we present an introduction to electroweak symmetry breaking 
and Higgs boson physics for the Weinberg-Salam model 
(the Standard Model \footnote{By Standard Model we understand the 
electroweak Weinberg-Salam model plus quantum chromodynamics.}).  
The current  experimental status of the Higgs boson searches and implications 
for future experiments at the Large Hadron Collider (LHC) are discussed. 
We don't review the Higgs boson physics at $e^+e^-$  linear collider 
\cite{7} and at muon collider \cite{8} because at present it is too far 
from reality.
It should be noted that  at present common belief is  that 
the Standard Model is not the whole story and at the $TeV$ scale 
new physics beyond Standard 
Model exists. Namely, the most popular scenario is the low energy broken 
supersymmetry with the $O(1)$ $TeV$ sparticle masses \cite{9}. 
In such scenario at least two Higgs boson doublets must exist, so in 
addition to the standard (light) Higgs boson $h$ there must exist scalar 
charged Higgs boson $H^{\pm}$, second 
neutral scalar Higgs boson $H$ and axial scalar Higgs boson $A$. For the most 
interesting case when the  Higgs boson $h$ is much lighter than the 
additional Higgs bosons $H, H^{\pm}, A$ we 
have the decoupling of the Heavy Higgs bosons  and 
the interactions of the lightest 
Higgs boson with vector bosons and fermions coincide up to power 
corrections with the Standard Model interactions. Therefore even if new 
physics beyond Standard Model exists at TeV region 
with very big probability the physics of the 
lightest Higgs boson is described by the Standard Model. Note that 
there are several books and reviews on the Higss boson 
physics \cite{10} - \cite{18}. 
The peculiarity of this review is that we give both theoretical aspects of 
the Higgs boson physics and experimental aspects related  to 
the search for the Higgs boson at LHC.     

The organization of the paper is the following. In section 2 we describe 
the Lagrangian of the Standard Model. In section 3 we give the main formulae 
for the Higgs boson decay widths. In section 4  indirect bounds on the 
Higgs boson mass are discussed. LEP1 and LEP2 Higgs boson mass bounds 
are given  in section 5. The Higgs boson production mechanisms and the 
main formulae for the cross sections are described in section 6.
In section 7 we discuss the possibilities to discover Higgs boson at 
upgrated TEVATRON. In section 8 we give review of the two main general 
purpose detectors at the LHC (CMS and ATLAS). The perspectives for 
the search for Higgs boson at the LHC are described in section 9.   
Section 10 contains concluding remarks.

\section{The Lagrangian of the  Standard Model}

The Standard  Model (SM) 
 is the renormalizable model of strong
and electroweak interactions. It has the gauge group
$SU(3)_c \otimes SU(2)_L \otimes U(1)$ and the minimal Higgs
structure consisting of one complex doublet of scalar particles. The
spontaneous electroweak symmetry breaking $SU(3)_c \otimes SU(2)_L
\otimes U(1) \rightarrow SU(3)_c \otimes U(1)_{EM}$ due to nonzero vacuum
expectation value of the Higgs doublet provides the simplest
realization of the Higgs mechanism \cite{5} which generates masses for
$W^{\pm}$, $Z$ gauge bosons and masses to quarks and leptons.  In
this approach, the Goldstone bosons are generated by dynamics of
elementary scalar fields and precisely one neutral Higgs scalar (the
Higgs boson) remains in the physical spectrum. The Lagrangian of
the Standard Model consists of several pieces:
\begin{equation}
L_{WS} = L_{YM} + L_{HYM} + L_{SH} + L_{f} + L_{Yuk}\,.
\end{equation}
Here $L_{YM}$ is the Yang-Mills Lagrangian without matter fields
\begin{equation}
L_{YM} =
-\frac{1}{4}F^i_{\mu\nu}(W)F^{\mu\nu}_i(W) - \frac{1}{4}
F^{\mu\nu}(W^0)F_{\mu\nu}(W^0) -
\frac{1}{4}F^a_{\mu\nu}(G)F_a^{\mu\nu}(G)\,,
\end{equation}
where $F^i_{\mu\nu}(W)$, $F^a_{\mu\nu}(G)$, $F_{\mu\nu}(W^0)$ are
given by
\begin{equation}
F^i_{\mu\nu}(W) = \partial_{\mu} W^i_{\nu} - \partial_{\nu}W^i_{\mu}
+g_2\epsilon^{ijk}W^j_{\mu}W^k_{\nu}\,,
\end{equation}
\begin{equation}
F_{\mu\nu}(W^0) = \partial_{\mu}W^0_{\nu} -
\partial_{\nu}W^{0}_{\mu}\,,
\end{equation}
\begin{equation}
F^{a}_{\mu\nu}(G) = \partial_{\mu}G^a_{\nu} - \partial_{\nu}G^a_{\mu}
+g_sf^{abc}G^b_{\mu}G^c_{\nu}\,,
\end{equation}
where $W^i_{\mu}$, $W^0_{\mu}$ are the $SU(2)_L \otimes U(1)$ gauge
fields, $G^a_{\mu}$ are the gluon fields and $\epsilon^{ijk}$,
$f^{abc}$ are the structure constants of the $SU(2)_L$ and $SU(3)_c$
gauge groups. The Lagrangian $L_{HYM}$ describes the Higgs doublet
interaction with $SU(2)_L \otimes U(1)$ gauge fields
\begin{equation}
L_{HYM} = (D_{L\mu}H)^{+}(D^{\mu}_LH)\,,
\end{equation}
where covariant derivatives are given by
\begin{equation}
D_{L\mu} = \partial_{\mu} -ig_1\frac{Y}{2}W^0_{\mu} - ig_2
\frac{\sigma^{i}}{2}W^i_{\mu}\,,
\end{equation}
\begin{equation}
D_{R\mu} = \partial_{\mu} -ig_1\frac{Y}{2}W^0_{\mu}\,,
\end{equation}
\begin{equation}
D^q_{L\mu} = \partial_{\mu} - ig_1\frac{Y}{2}W^0_{\mu} - ig_2
\frac{\sigma^{i}}{2}W^{i}_{\mu} - ig_3t^aG^a_{\mu}\,,
\end{equation}
\begin{equation}
D^q_{R\mu} = \partial_{\mu} - ig_1\frac{Y}{2}W^0_{\mu} -
ig_3t^aG^{a}_{\mu}\,.
\end{equation}
Here $g_1$ is the $U(1)$ gauge coupling constant, 
$g_2$ and $g_3$ are  the $SU(2)_L$ and $SU(3)_c$ gauge coupling 
constants, $Y$ is the hypercharge
determined by the relation $Q = \frac{\sigma_{3}}{2} + \frac{Y}{2}$,
$\sigma^{i}$ are the Pauli matrices, $t^{a}$ are $SU(3)$ matrices in the
fundamental representation,
$H = \left( \begin{array}{cc}
H_1\\
H_2
\end{array}\right)$
is the Higgs $SU(2)_L$ doublet with $Y = 1$. The
Lagrangian $L_{SH}$ describing Higgs doublet self-interaction has the
form
\begin{equation}
L_{SH} = -V_0(H) = M^2H^{+}H - \frac{\lambda}{2}(H^{+}H)^2\,,
\end{equation}
where $H^{+}H = \sum_{i}H^{*}_iH_i$ and $\lambda$ is the Higgs
self-coupling constant. The Lagrangian $L_{f}$ describes the interaction
of fermions with gauge fields. Fermions constitute only doublets and
singlets in $SU(2)_L \otimes U(1)$
\begin{equation}
R_1 = e_R,\;R_2 =\mu_{R},\;R_{3} = \tau_{R} \,,
\end{equation}
\begin{equation}
L_1 = {\nu \choose e}_L \; L_2 ={\nu^{'} \choose \mu}_L \;
L_3 = {\nu^{''} \choose \tau}_L\,
\end{equation}
\begin{equation}
R_{qIu} = (q_{Iu})_R, \;\; (q_{1u} = u, \; q_{2u} = c, \; q_{3u} = t)\,,
\end{equation}
\begin{equation}
R_{qid} =(q_{id})_R, \;\; (q_{1d} = d, \; q_{2d} = s, \; q_{3d} =b)\,,
\end{equation}
\begin{equation}
L_{qI} = {q_{Iu} \choose V_{Ii}q_{id}}_L \,,
\end{equation}
where L and R denote left- and right-handed components of the
spinors respectively,
\begin{equation}
\psi_{R,L} = \frac{1 \pm \gamma_{5}}{2} \psi
\end{equation}
and $V_{Ii}$ is the Kobayashi-Maskawa matrix. The neutrinos are
assumed to be left-handed and massless. The Lagrangian $L_{f}$ 
describes the interaction of fermions with gauge fields and it has
the form
\begin{equation}
L_{f} =\sum_{k = 1}^{3}[ i\bar{L}_{k}\hat{D}_{L}L_{k} +
i\bar{R}_{k}\hat{D}_{R}R_{k} + i\bar{L}_{qk}\hat{D}^q_LL_{qk}
+ i\bar{R}_{qku}\hat{D}^q_RR_{qku} +
i\bar{R}_{qkd}\hat{D}^q_RR_{qkd}] \,,
\end{equation}
where $\hat{D}_L = \gamma^{\mu}D_{L\mu}$, $\hat{D}_{R} =
\gamma^{\mu}D_{R\mu}$, $\hat{D}^q_L = \gamma^{\mu}D^q_{L\mu}$,
$\hat{D}^q_R = \gamma^{\mu}D^q_{R\mu}$.
The Lagrangian $L_{Yuk}$ generates fermion mass terms. Supposing the
neutrinos to be massless,  the Yukawa interaction of the
fermions with Higgs doublet has  the form
\begin{equation}
L_{Yuk} = -\sum_{k=1}^{3}[h_{lk}\bar{L}_kHR_k + h_{dk}\bar{L}_{qk}^{'}H
R_{dk} + h_{uk}\bar{L}_{qk}^{'}(i\sigma^{2}H^{*})R_{uk}] \,+ h.c.\,,
\end{equation}
$$ L_{qI}^{'} = {q_{Iu} \choose q_{Id}} $$.
The potential term $V_0(H)= -M^2H^{+}H  +
\frac{\lambda}{2}(H^{+}H)^2$
for $M^2 > 0$  gives rise to the
spontaneous  symmetry breaking. The doublet $H$ acquires the nonzero
vacuum expectation value
\begin{equation}
<H> = \left( \begin{array}{cc}
0\\
\frac{v}{\sqrt{2}}
\end{array} \right)\,,
\end{equation}
where $v = 246 $ GeV. In the unitare gauge unphysical Goldstone
massless fields are absent and the Higgs doublet scalar field depends
on the single physical scalar field $h(x)$ (Higgs boson field):
\begin{equation}
H(x) = \left( \begin{array}{cc}
0\\
\frac{v}{\sqrt{2}} + \frac{h(x)}{\sqrt{2}}
\end{array} \right)\,.
\end{equation}
Due to spontaneous gauge symmetry breaking gauge fields except gluon and 
photon fields acquire masses. Diagonalization of mass matrix gives
\begin{equation}
W_{\mu}^{\pm} = \frac{1}{\sqrt{2}}(W^1_{\mu} \mp W^2_{\mu}), \; M_W =
\frac{1}{2}g_2v \,,
\end{equation}
\begin{equation}
Z_{\mu} = \frac{1}{\sqrt{g^2_2 + g^2_1}}(g_2W^3_{\mu} -g_1W^0_{\mu}),
\; M_Z = \frac{1}{2}\sqrt{g^2_2 +g^2_1}\,v\,,
\end{equation}
\begin{equation}
A_{\mu} = \frac{1}{\sqrt{q^2_2 + g^2_1}}(g_1W^3_{\mu} + g_2W^0_{\mu}),
\; M_A = 0 \,,
\end{equation}
where $W^{\pm}_{\mu}$, $Z_{\mu}$ are charged and neutral 
electroweak boson fields, $A_{\mu}$ is photon field. It is convenient 
to introduce rotation angle
$\theta_{W}$ between $(W^3,W^0)$ and $(Z,A)$ which is called Weinberg angle
\begin{equation}
\sin{\theta_{W}} \equiv \frac{g_1}{\sqrt{g^2_1 + g^2_2}} \,.
\end{equation}
Experimentally $\sin^{2}{\theta_{W}} \approx 0.23$ \cite{19}.
The formula for the electric charge $e$ has the form
\begin{equation}
e = \frac{g_2g_1}{\sqrt{g^2_2 + g^2_1}}\,.
\end{equation}
At the tree level the Higgs boson mass is determined by the
formula
\begin{equation}
m_h = \sqrt{2} M = \sqrt{\lambda} v \,.
\end{equation}
The Lagrangian $L_{HYM}$ describes the interaction of the 
Higgs boson field with vector $W$- and $Z$-bosons. In the unitare gauge 
it reads
\begin{equation}
L_{HYM} = \frac{1}{2}\partial^{\mu}h \partial_{\mu}h +
M^2_W(1 + \frac{h}{v})^2 W^{+}_{\mu}W^{\mu} + \frac{1}{2}
M^2_Z(1 + \frac{h}{v})^2Z^{\mu}Z_{\mu} \,.
\end{equation}
The Lagrangian $L_{Yuk}$ is responsible for the fermion masses
generation. In the unitare gauge it can be written
in the form
\begin{equation}
L_{Yuk} = -\sum_{i} m_{\psi_{i}}(1 + \frac{h}{v})
 \bar{\psi}_{i} \psi_{i} \,,
\end{equation}
where $\psi_{i}$ are the fermion(quark and lepton) fields.

\section{Indirect Higgs boson mass bounds}

\subsection{Tree level unitarity}

The Higgs boson has been introduced as a fundamental particle to render 
2 - 2 scattering amplitudes (see Fig.1) involving longitudinally 
polarized W bosons 
compatible with unitarity. In general particles must decouple from low 
energy spectrum if their mass grows indefinitely. Therefore the Higgs 
boson mass must be bounded to restore unitarity in the perturbation 
theory. The asymptotic tree level formula for the elastic $W_LW_L$ 
 S-wave scattering amplitude reads \cite{20,21}
\begin{equation}
A^{J=0}(W_LW_L \rightarrow W_LW_L) \approx -\frac{G_Fm^2_h}{4\sqrt{2}\pi} .
\end{equation}
Partial wave unitarity implies that
\begin{equation}
|A^{J}|^2 \leq |Im(A^{J})|,
\end{equation}
\begin{equation}
(Re(A^{J}))^2 \leq |Im(A^J)(1 - |Im(A^J)|)|. 
\end{equation}
As a consequence we find that 
\begin{equation}
|Re(A^J)| \leq \frac{1}{2}
\end{equation}
Hence \cite{20,21},
\begin{equation}
m^2_h \leq \frac{2\pi\sqrt{2}}{G_F} \approx (850 ~GeV)^2
\end{equation}

The most stringent bound is obtained by performing a full coupled channel 
analysis for the scattering of longitudinal gauge bosons into 
$W^+_LW^-_L, Z_LZ_L, Z_Lh$ and $hh $. The largest eigenvalue of the amplitude 
matrix gives the most restrictive bound 
\begin{equation}
m^2_h \leq \frac{4\pi\sqrt{2}}{3G_F} \approx (700 ~GeV)^2
\end{equation}
However it should be noted that if $m_h \geq 700 ~GeV $ it means simply 
that perturbation theory is no longer reliable and in principle 
an account of higher order corrections can restore unitarity. Lattice 
estimates give similar bound \cite{22} $m_h \leq ~700 ~GeV$ on the 
Higgs boson mass.

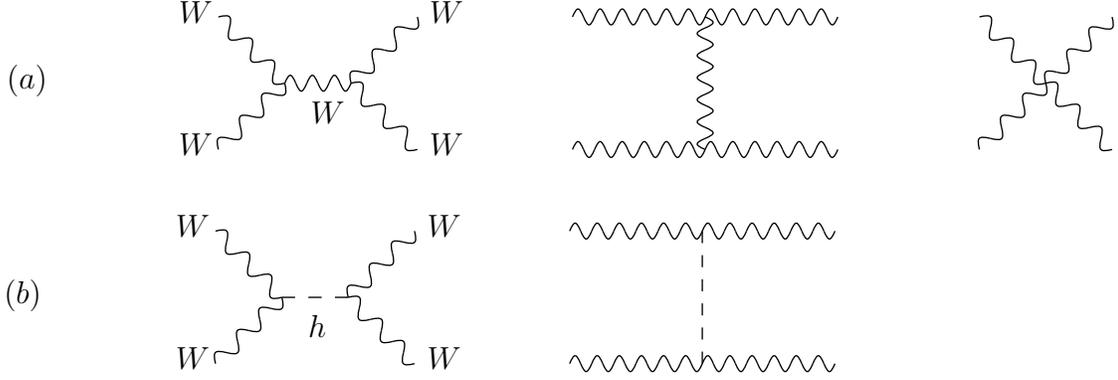
\begin{figure}[hbt]
\begin{center}
\begin{picture}(80,80)(50,-20)
\Photon(0,50)(25,25){3}{3}
\Photon(0,0)(25,25){3}{3}
\Photon(25,25)(50,25){3}{3}
\Photon(50,25)(75,50){3}{3}
\Photon(50,25)(75,0){3}{3}
\put(-80,23){$(a)$}
\put(-15,-2){$W$}
\put(-15,48){$W$}
\put(80,-2){$W$}
\put(80,48){$W$}
\put(35,10){$W$}
\end{picture}
\begin{picture}(60,80)(0,-20)
\Photon(0,50)(100,50){3}{12}
\Photon(0,0)(100,0){3}{12}
\Photon(50,50)(50,0){3}{6}
\end{picture}
\begin{picture}(60,80)(-90,-20)
\Photon(0,50)(25,25){3}{3}
\Photon(0,0)(25,25){3}{3}
\Photon(25,25)(50,50){3}{3}
\Photon(25,25)(50,0){3}{3}
\end{picture} \\
\begin{picture}(80,60)(83,0)
\Photon(0,50)(25,25){3}{3}
\Photon(0,0)(25,25){3}{3}
\DashLine(25,25)(50,25){6}
\Photon(50,25)(75,50){3}{3}
\Photon(50,25)(75,0){3}{3}
\put(-80,23){$(b)$}
\put(-15,-2){$W$}
\put(-15,48){$W$}
\put(80,-2){$W$}
\put(80,48){$W$}
\put(35,10){$h$}
\end{picture}
\begin{picture}(60,60)(33,0)
\Photon(0,50)(100,50){3}{12}
\Photon(0,0)(100,0){3}{12}
\DashLine(50,50)(50,0){5}
\end{picture}  \\
\end{center}
\caption[]{\it \label{fg:wwtoww} Tree level  diagrams of elastic 
$WW$ scattering:
(a) pure gauge-boson dynamics, and (b) Higgs-boson exchange.}
\end{figure}

\subsection{Vacuum stability bound}

It is possible also to derive bounds on the Higgs boson mass 
from the requirement  of the absence of the Landau pole singularity
for the effective Higgs self-coupling constant \cite{23} and from
the vacuum stability requirement \cite{24}.

The idea of the derivation of the bound resulting from the requirement of 
the absence of Landau pole singularities is the following \cite{23}. 
Suppose Standard Model is valid up to the scale $\Lambda$. We require that 
the effective Higgs self-coupling constant does not have Landau pole 
singularities up to the energies $\Lambda$. From this requirement we 
find an upper  bound on the low energy Higgs self-coupling 
constant $\bar{\lambda}(m_t)$ which determines the Higgs boson mass. 
Namely, the renormalization group equations for the effective coupling
constants in neglection of all Yukawa coupling constants except
top-quark Yukawa coupling constant
in one-loop approximation read
\begin{equation}
\frac{d\bar{g}_3}{dt} = -7\bar{g}^3_3\,,
\end{equation}
\begin{equation}
\frac{d\bar{g}_2}{dt} = -(\frac{19}{6})\bar{g}^3_2\,,
\end{equation}
\begin{equation}
\frac{d\bar{g}_1}{dt} = (\frac{41}{6})\bar{g}^3_1\,,
\end{equation}
\begin{equation}
\frac{d\bar{h}_t}{dt} = (\frac{9\bar{h}^2_t}{2} -8\bar{g}^2_3 -
\frac{9\bar{g}^2_2}{4} -\frac{17\bar{g}^2_1}{12})\bar{h}_t\,,
\end{equation}
\begin{equation}
\frac{d\bar{\lambda}}{dt} = 12(\bar{\lambda}^2 + (\bar{h}^2_t -
\frac{\bar{g}^2_1}{4} - \frac{3\bar{g}^2_2}{4})\lambda -
\bar{h}^4_t
+ \frac{\bar{g}^4_1}{16} + \frac{\bar{g}^2_1\bar{g}^2_2}{8} +
\frac{3\bar{g}^4_2}{16})\,,
\end{equation}
\begin{equation}
t = (\frac{1}{16\pi^2})\ln{(\mu/m_Z)}\,.
\end{equation}
Here $\bar{g}_3$, $\bar{g}_2$ and $\bar{g}_1$ are the $SU(3)_c$,
$SU(2)_L$ and $U(1)$ effective gauge couplings, respectively, 
and $\bar{h}_t$ is the effective top quark Yukawa coupling constant. 
In our concrete estimates we
took $m_t^{pole} = 175$ GeV, $\bar{\alpha}_3(m_Z) = 0.118$,
$\bar{\alpha}_{em}^{-1}(m_Z) = 127.9$, $\sin^{2}{\theta_W}(m_Z) =
0.2337$, $\alpha_i \equiv \frac{g^2_i}{4\pi}$. From the requirement
of the absence of Landau pole singularity for the Higgs self-coupling
constant $\bar{\lambda}$ for the scales  up to $\Lambda = (10^{3}; 10^{4};
10^{6}; 10^{8}; 10^{10}; 10^{12}; 10^{14})$ GeV
(to be precise we require that at the scale $\Lambda$ the Higgs
self-coupling constant is $\frac{\bar{\lambda}^{2}(\Lambda)}{4\pi} \leq
1$) we have found the upper bound on the Higgs boson mass $m_h \leq
(400; 300; 240; 200; 180; 170; 160)$ GeV, respectively. 

The vacuum stability bound \cite{24} comes from the requirement that 
the electroweak minimum of the effective potential is the deepest one 
for $|H| \leq \Lambda$. Remember that $\Lambda$ is the scale up 
to which the Standard Model is assumed to be valid. For $|H| \gg v$ 
the mass terms in the effective potential are negligible compared to 
the self-interaction term and the vacuum stability requirement 
 means that  
the Higgs self-interaction coupling is nonegative
$\bar{\lambda}(\mu) \geq 0$ for the scales $\mu \leq \Lambda $. 
Suppose that at scales $M \geq M_s$ we have some supersymmetric 
extension of the Standard Model. It should be noted that 
the most popular at present is the minimal supersymmetric 
standard model (MSSM) \cite{9} which predicts that
the effective Higgs self-coupling constant for the standard model
at the scale  of supersymmetry breaking $M_s \equiv \Lambda $ has to obey the
inequality
\begin{equation}
0 \leq \bar{\lambda}(M_s) = (\bar{g}^2_1(M_s) + \bar{g}^2_2(M_s))
(\cos(2\varphi))^2/4 \leq (\bar{g}^2_1(M_s) + \bar{g}^2_2(M_s))/4 \,.
\end{equation}
So the assumption that standard Weinberg-Salam model originates
from its supersymmetric extension with the supersymmetry broken
at scale $M_s$ allows us to obtain non-trivial information about
the low energy effective Higgs self-coupling constant in the
effective potential $V = - M^2H^{+}H  +
\frac{\lambda}{2}(H^{+}H)^2$ and hence to obtain nontrivial
information about the Higgs boson mass. It should be noted that
in nonminimal supersymmetric electroweak models, say in the model
with additional gauge  singlet $\sigma$, we have due to the
$k\sigma H_1 i \tau_{2}H_2$ term in the superpotential an additional
term $k^2|H_1i\tau_{2}H_{2}|^2$ in the potential and as a consequence
our boundary condition for the Higgs self-coupling constant has to be
modified, namely
\begin{equation}
\bar{\lambda}(M_s) = \frac{1}{4}(\bar{g}^2_1(M_s) + \bar{g}^2_2(M_s))
\cos^2(2\varphi)
+ \frac{1}{2}\bar{k}^2(M_s)\sin^{2}(2\varphi) \geq 0 \,.
\end{equation}
The boundary condition  (43) depends on unknown coupling constant
$\bar{k}^2(M_s)$. However it is very important to stress that for all
nonminimal supersymmetric models broken to standard Weinberg-Salam
model at scale $M_s$ the effective Higgs self-coupling constant
$\bar{\lambda}(M_s)$  is  non-negative which is a direct
consequence of the non-negativity of the effective potential
in supersymmetric models. Therefore the vacuum stability
requirement results naturally \cite{25} if supersymmetry is broken at
some high scale $M_s$ and at lower scales the  Weinberg-Salam
model is an effective theory. For the Weinberg-Salam model 
 with boundary condition
(42) for the Higgs self-coupling constant $\bar{\lambda}(M_s)$
we have integrated numerically renormalization group equations
in two-loop approximation. Also we took into account the
one loop correction to the Higgs boson mass
 (running Higgs boson mass $\bar{m}_h(\mu) =
\sqrt{\bar{\lambda}(\mu)}v$ does not coincide with pole
Higgs boson mass). Our results \cite{25} for the Higgs boson mass 
$m_h(k,M_s,m_t^{pole})$ for different values of $M_s$ and  
$m_t^{pole}$ are presented in
table 1. Here $k = 0$ corresponds to the boundary condition
$\bar{\lambda}(M_s) = 0$ (vacuum stability bound) and 
$k = 1$ corresponds to
the boundary condition $\bar{\lambda}(M_s) = \frac{1}{4}
(\bar{g}^2_1 + \bar{g}^2_2)$. So  from the requirement 
that at some high scale $M_s$ the MSSM is softly broken to the SM 
we find \cite{25} that the Higgs boson mass lies in the interval 
$$ m_h(k=0,M_s,m_t^{pole}) \leq m_h \leq m_h(k=1,M_s,m_t^{pole}) $$. 
The accuracy in the determination of $m_h(k,M_s,m_t^{pole})$ is 
related mainly to nonexact knowledge of $\alpha_3(M_Z)$ and it is 
estimated to be less than $3~GeV$. For instance, for $m_t^{pole} = 175~GeV$ 
and $M_s = 10^{8}~GeV$ we find that
$$129~GeV \leq m_h \leq 147~GeV$$.

Table 1. The dependence of the Higgs boson mass 
$m_h(k,M_s,m_t^{pole})$ on the values of
$M_s$, $m_t^{pole}$ and $k =0,1$. Everything except $k$ is in $GeV$.

\begin{center}
\begin{tabular}{|l| |l| |l| |l| |l| |l| |l| |l| |l| |l| |l|}
\hline
$m_t^{pole}$ & 165 & 165 & 170 & 170 &175 & 175 & 180 & 180 & 185 & 185 \\
\hline
 &k=0 &k=1& k=0 & k=1 &k=0& k=1 &k=0& k=1& k=0& k=1\\
\hline
$M_s=10^{3}$&69&111&74&114&78&117&83&120&88&123\\
\hline
$M_s=10^{3.5}$&81&117&86&120&92&124&98&128&104&132\\
\hline
$M_s=10^{4}$&89&121&95&125&101&130&108&134&114&139\\
\hline
$M_s=10^{6}$&105&129&113&135&121&141&129&147&137&153\\
\hline
$M_s=10^{8}$&112&132&120&138&129&147&138&152&146&159\\
\hline
$M_s=10^{10}$&115&133&124&140&133&147&142&154&151&161\\
\hline
$M_s=10^{12}$&117&134&126&141&136&147&145&154&154&161\\
\hline
$M_s=10^{14}$&118&134&127&141&132&148&147&156&156&164\\
\hline
$M_s=10^{16}$&118&134&128&141&138&148&148&156&158&164\\
\hline
\end{tabular}
\end{center}

Note that in the MSSM the mass of the lightest Higgs boson is 
less than $m_h \leq M_Z$ at tree level. Radiative corrections 
can increase the mass of the lightest Higgs boson \cite{26} up 
to $120 ~GeV$ provided  the sparticle masses are less than $1~TeV$. 
As it has been demonstrated in refs. \cite{27} in Standard Model 
due to the vacuum stability condition the Higgs boson mass 
has to be heavier than $\sim 120 ~GeV$ \footnote{Concrete details 
and rigorous statements are contained in refs.\cite{27}}.  
It means that by the measurement of the Higgs boson mass it 
would be possible to distinguish between SM and MSSM. In 
particular, the observation of the Higgs boson at LEP2 with 
a mass less than $110 ~GeV$ will be powerful untrivial indication 
in favour of the existence of low energy broken supersymmetry.

\subsection{Higgs boson mass bound from electroweak precision data}

Indirect bound on the Higgs boson mass can be derived from the high-precision 
measurements of electroweak observables at LEP and elsewhere. The 
Standard Model is renormalizable only after including the top quark and 
the Higgs boson and as a consequence the electroweak observables are 
sensitive to the masses of these particles. The Fermi coupling can be 
rewritten as 
\begin{equation}
\frac{G_F}{\sqrt{2}} = \frac{2\pi\alpha}{\sin^2(2\theta_W)M^2_Z}
[1 + \Delta r_a + \Delta r_t + \Delta r_h],
\end{equation}
The $\Delta$ terms take into account the radiative corrections: 
$\Delta r_a$ describes the shift in the effective electromagnetic 
coupling constant; $\Delta r_t$ takes into account the top quark contribution.
 The $\Delta r_ h$ denotes the Higgs boson contribution. This term depends 
logarithmically \cite{28} on the Higgs boson mass and at leading 
order it reads 
\begin{equation}
\Delta r_h = \frac{11G_FM^2_W}{24 \sqrt{2} \pi}[log(\frac{m^2_h}{M^2_W}) 
-\frac{5}{6}] ,\,\,  (m^2_h \gg M^2_W)
\end{equation}
 Although the sensitivity on the Higgs boson mass is only 
logarithmic, the increasing precision in the measurement of the electroweak 
observables allows to derive constraints on the Higgs boson mass 
\cite{29}
\begin{equation}
m_h = 71 ^{+75}_{-42} \pm 5 ~GeV. 
\end{equation} 
In other words it means that the Higgs boson should be relatively light 
with a mass less than $m_h <220 ~GeV$ at $95 \,\% \, C.L.$ \cite{29}.
See however, ref. \cite{30} where 
it has been shown on the base of the scale factor fit that 95 percent 
 confidence level upper limit increases to as much as $750~GeV$. 

\section{Higgs boson decays}

The tree-level Higgs boson couplings to gauge bosons and fermions
can be deduced from the Lagrangian (28, 29). Of these, the
$hW^+W^-$, $hZZ$ and $h\bar{\psi} \psi$ are the most important for
the phenomenology. The partial decay width into fermion-antifermion
pair is \cite{31} 
\begin{equation}
\Gamma(h \rightarrow \psi \bar{\psi}) = \frac{G_Fm^2_{\psi}m_hN_c}
{4\pi \sqrt{2}}(1 -\frac{4m^2_{\psi}}{m^2_h})^{\frac{3}{2}}\,,
\end{equation}
where $N_c$ is the number of fermion colours. For $m_{h} \leq 2m_W$
Higgs boson decays mainly with ($\approx$ 90 percent) probability
into b quark-antiquark pair and with $\approx$ 5 percent probability
into $\tau$ lepton-antilepton pair. An account of higher order QCD
corrections can be effectively taken into account in the formula
(47) for the Higgs boson decay into b quark-antiquark pair by the
replacement of pole b-quark mass in formula (47) by the effective
b-quark mass $\bar{m}_b(m_h)$.
An account of higher order corrections leads to the formula \cite{32} 
(see Fig.2) 
\begin{equation}
\Gamma(h \rightarrow Q\bar{Q}) = \frac{3G^2_Fm_h}{4\sqrt{2}\pi}
\bar{m}^2_Q(m_h)[\Delta_{QCD} +\Delta_t],
\end{equation}
\begin{eqnarray}
&&\Delta_{QCD} = 1 + 5.67 \frac{\alpha_s(m_h)}{\pi} +
(35.94 - 1.36N_F)(\frac{\alpha_s(m_h)}{\pi})^2 + \\ \nonumber
&&(161.14 -25.77N_F +0.259N^2_F)(\frac{\alpha_s(m_h)}{\pi})^3,
\end{eqnarray}
\begin{equation}
\Delta_t = (\frac{\alpha_s(m_h)}{\pi})^2[1.57 -\frac{2}{3}
\log\frac{m^2_h}{m^2_t} +\frac{1}{9} \log^2\frac{\bar{m}^2_Q(m_h)}{m^2_h}]
\end{equation}
for the Higgs boson decay width to $Q = b,c$ quarks in the $\bar{MS}$ 
renormalization scheme. The relation between the perturbative quark pole mass 
$m_Q$ and the $\bar{MS}$ running quark mass $\bar{m}_Q(m_Q)$
has the form \cite{33}
\begin{equation}
\bar{m}_Q(m_Q) = \frac{m_Q}
{1 +\frac{4}{3}\frac{\alpha_s(m_Q)}{\pi} + K_Q
(\frac{\alpha_s(m_Q)}{\pi})^2},
\end{equation}
where numerically $K_t \approx 10.9$, $K_b \approx 12.4$ and 
$K_c \approx 13.4$. Electroweak corrections to heavy quarks and lepton decays 
are rather small \cite{34}(less than 2 percent).

\begin{figure}[hbt]
\begin{center}
\setlength{\unitlength}{1pt}
\begin{picture}(350,300)(0,0)


\ArrowLine(30,250)(60,280)
\ArrowLine(60,220)(30,250)
\DashLine(0,250)(30,250){5}
\put(-15,246){$h$}
\put(65,280){$Q$}
\put(65,213){$\bar Q$}


\Gluon(175,235)(175,265){-3}{3}
\ArrowLine(160,250)(175,265)
\ArrowLine(175,265)(190,280)
\ArrowLine(190,220)(175,235)
\ArrowLine(175,235)(160,250)
\DashLine(130,250)(160,250){5}
\put(115,246){$h$}
\put(195,280){$Q$}
\put(195,213){$\bar Q$}
\put(185,250){$g$}

\Gluon(285,235)(310,235){3}{2}
\ArrowLine(270,250)(300,280)
\ArrowLine(300,220)(285,235)
\ArrowLine(285,235)(270,250)
\DashLine(240,250)(270,250){5}
\put(225,246){$h$}
\put(305,280){$Q$}
\put(305,213){$\bar Q$}
\put(315,233){$g$}


\Gluon(40,140)(40,160){-3}{2}
\Gluon(50,130)(50,170){-3}{4}
\ArrowLine(30,150)(40,160)
\ArrowLine(40,160)(50,170)
\ArrowLine(50,170)(60,180)
\ArrowLine(60,120)(50,130)
\ArrowLine(50,130)(40,140)
\ArrowLine(40,140)(30,150)
\DashLine(0,150)(30,150){5}
\put(-15,146){$h$}
\put(65,180){$Q$}
\put(65,113){$\bar Q$}
\put(60,150){$g$}

\Gluon(155,135)(170,135){3}{1}
\Gluon(170,135)(185,150){3}{2}
\Gluon(170,135)(185,120){-3}{2}
\ArrowLine(140,150)(170,180)
\ArrowLine(170,120)(155,135)
\ArrowLine(155,135)(140,150)
\DashLine(110,150)(140,150){5}
\put(95,146){$h$}
\put(175,180){$Q$}
\put(171,107){$\bar Q$}
\put(190,120){$g$}
\put(190,148){$g$}

\DashLine(230,150)(260,150){5}
\ArrowLine(260,150)(290,180)
\ArrowLine(290,120)(260,150)
\ArrowLine(290,180)(290,120)
\Gluon(290,180)(320,180){3}{3}
\Gluon(290,120)(320,120){-3}{3}
\ArrowLine(320,120)(320,180)
\ArrowLine(320,180)(350,180)
\ArrowLine(350,120)(320,120)
\put(215,146){$h$}
\put(355,180){$Q$}
\put(355,113){$\bar Q$}
\put(260,160){$t$}
\put(305,170){$g$}
\put(305,130){$g$}


\Gluon(50,40)(57,50){-3}{1}
\Gluon(57,50)(50,60){-3}{1}
\Gluon(70,30)(57,50){-3}{2}
\Gluon(57,50)(70,70){-3}{2}
\ArrowLine(30,50)(50,60)
\ArrowLine(50,60)(70,70)
\ArrowLine(70,70)(80,75)
\ArrowLine(80,25)(70,30)
\ArrowLine(70,30)(50,40)
\ArrowLine(50,40)(30,50)
\DashLine(0,50)(30,50){5}
\put(-15,46){$h$}
\put(85,80){$Q$}
\put(85,13){$\bar Q$}
\put(70,50){$g$}

\Gluon(175,35)(175,50){-3}{1}
\Gluon(175,50)(175,65){-3}{1}
\Gluon(175,50)(200,65){-3}{2}
\Gluon(175,50)(200,35){3}{2}
\ArrowLine(160,50)(175,65)
\ArrowLine(175,65)(190,80)
\ArrowLine(190,20)(175,35)
\ArrowLine(175,35)(160,50)
\DashLine(130,50)(160,50){5}
\put(115,46){$h$}
\put(195,80){$Q$}
\put(195,13){$\bar Q$}
\put(205,63){$g$}
\put(205,33){$g$}

\Gluon(295,65)(325,65){3}{3}
\Gluon(295,35)(310,35){3}{1}
\Gluon(310,35)(325,50){3}{2}
\Gluon(310,35)(325,20){-3}{2}
\ArrowLine(280,50)(295,65)
\ArrowLine(295,65)(310,80)
\ArrowLine(310,20)(295,35)
\ArrowLine(295,35)(280,50)
\DashLine(250,50)(280,50){5}
\put(235,46){$h$}
\put(315,80){$Q$}
\put(311,07){$\bar Q$}
\put(330,20){$g$}
\put(330,48){$g$}
\put(330,63){$g$}

\end{picture}  \\
\setlength{\unitlength}{1pt}
\caption[ ]{\label{fg:hqqdia} \it Typical diagrams contributing to $h \to
Q\bar Q$ at lowest order and one-, two- and three-loop QCD.}
\end{center}
\end{figure}
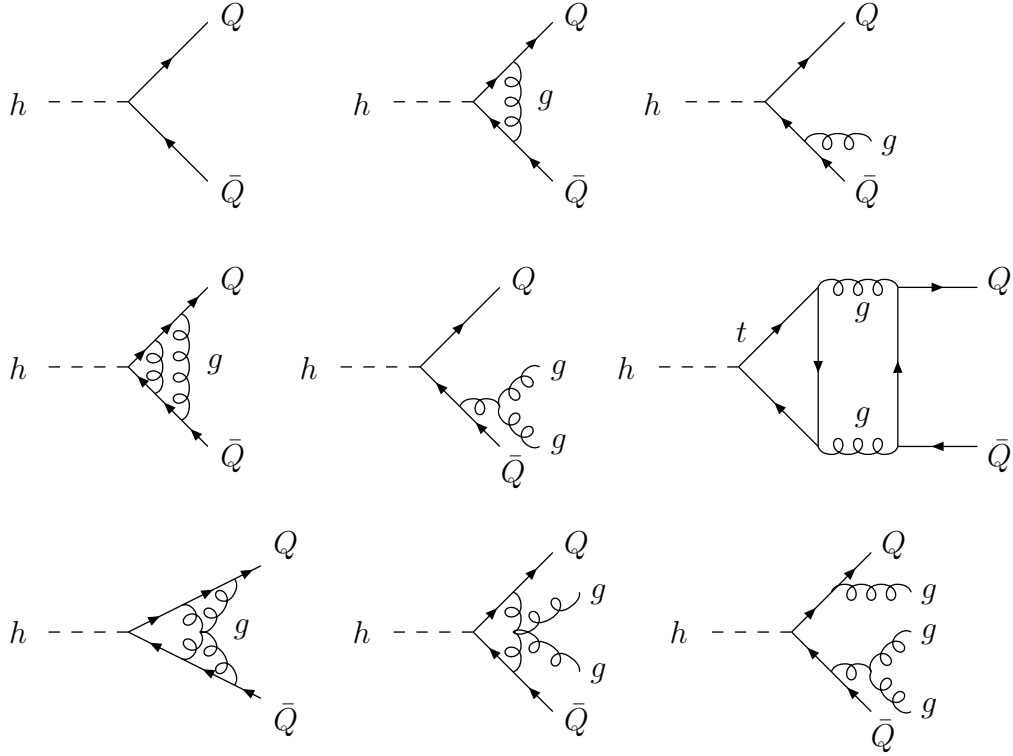

Higgs boson with $m_h \geq 2M_{W}$ will decay into pairs of
gauge bosons (see Fig.3) with the partial widths
\begin{equation}
\Gamma(h \rightarrow W^{+}W^{-}) = \frac{G_Fm^3_h}{32\pi\sqrt{2}}
(4 -4a_w +3a^2_w)(1-a_w)^{\frac{1}{2}} \,,
\end{equation}
\begin{equation}
\Gamma(h \rightarrow Z^0 Z^0) = \frac{G_Fm^3_h}{64\pi\sqrt{2}}
(4 - 4a_Z + 3a^2_Z)(1 -a_Z) ^{\frac{1}{2}} \,,
\end{equation}
where $a_W = \frac{4M^2_W}{m^2_h}$ and $a_Z=\frac{4M^2_Z}{m^2_h}$.
The electroweak corrections have been computed in refs. \cite{34}.
They are less than 5 percent in the intermediate region. The QCD 
corrections to the leading top mass corrections of $O(G_Fm^2_t)$ have 
been calculated in refs. \cite{35}.

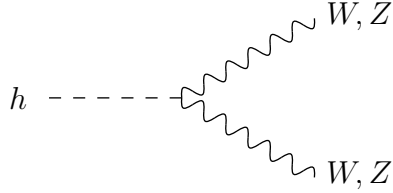
\begin{figure}[hbt]
\begin{center}
\setlength{\unitlength}{1pt}
\begin{picture}(200,100)(0,0)

\DashLine(0,50)(50,50){5}
\Photon(50,50)(100,80){3}{6}
\Photon(50,50)(100,20){-3}{6}
\put(-15,46){$h$}
\put(105,18){$W,Z$}
\put(105,78){$W,Z$}

\end{picture}  \\
\setlength{\unitlength}{1pt}
\caption[ ]{\label{fg:hvvdia} \it Diagram contributing to $h\to VV$ [$V=W,Z$].}
\end{center}
\end{figure}

In the heavy Higgs mass regime $(2m_Z \leq m_h \leq 800$ GeV), the
Higgs boson decays dominantly into gauge bosons. For example,
for $m_h \gg 2m_Z$ one can find that
\begin{equation}
\Gamma(h \rightarrow W^{+}W^{-}) \simeq  2\Gamma(h \rightarrow ZZ)
\simeq \frac{G_Fm^3_h}{8\pi \sqrt{2}}\,.
\end{equation}
The $m^3_h$ behaviour is a consequence of the longitudinal
polarisation states of the $W$ and $Z$. As $m_h$ gets large, so does
the coupling of $h$ to the Goldstone bosons which have been eaten by
the $W$ and $Z$. However, the Higgs boson decay width to a pair of
heavy quarks growth only linearly in the Higgs boson mass. Thus, for
the Higgs masses sufficiently above $2m_Z$, the total Higgs boson
width is well approximated by ignoring the Higgs boson decay to $t\bar{t}$
and including only the two gauge boson modes. For heavy Higgs
boson mass one can find that
\begin{equation}
\Gamma_{total}(h) \simeq 0.48\,TeV(\frac{m_h}{1\,TeV})^3 \,.
\end{equation}
For large Higgs boson mass higher order corrections due to 
the self-coupling of the Higgs boson are relevant, namely \cite{36}
\begin{equation}
\Gamma(h \rightarrow VV) = \Gamma_{LO}(h \rightarrow VV)[1 + 2.8 \kappa + 
62.0 (\kappa)^2],
\end{equation}
where $\kappa = \frac{G_Fm^2_h}{16\sqrt{2}\pi^2}$, $V = Z,W$.

Below threshold the decays into off-shell gauge particles are important. 
The decay width  into single off-shell gauge boson has the form \cite{37}
\begin{equation}
\Gamma(h \rightarrow V V^*) = \delta_V \frac{3G^2_FM^4_Vm_h}{16\pi^3}
R(\frac{M^2_V}{m^2_h}),
\end{equation}
where $\delta_W =1 $, $\delta_Z = \frac{7}{12} -\frac{10}{9}\sin^2\theta_W 
+ \frac{40}{27}\sin^4\theta_W$ and 
\begin{equation}
R(x) = 3 \frac{1-8x+20x^2}{\sqrt{4x-1}}\arccos(\frac{3x-1}{2x^{3/2}}) -
\frac{1-x}{2x}(2 -13x +47x^2) -\frac{3}{2}(1 -6x + 4x^2)\log(x),
\end{equation}
$x = \frac{M^2_V}{m^2_h}$. For Higgs boson mass slightly larger than the 
corresponding gauge boson mass the decay widths into pairs of off-shell 
gauge bosons play important role. The corresponding formulae can be found in 
ref. \cite{38}. 

It should be noted that there are a number of important Higgs boson couplings
which are absent at tree level but appear at one-loop level.
Among them the couplings of the Higgs boson  to two
gluons and two photons are extremely important for the Higgs
boson searches at supercolliders. One-loop induced Higgs
coupling to two gluons is due to t-quark exchange in the loop (see Fig.4)
\cite{39} and it leads to an effective Lagrangian
\begin{equation}
L^{eff}_{hgg} = \frac{g_2\alpha_s}{24\pi m_W}hG^a_{\mu\nu}
G^{a\mu\nu} \,.
\end{equation}
for the interaction of the Higgs boson with gluons.
At lowest order the partial decay width is given by \cite{39}
\begin{equation}
\Gamma_{LO}(h \rightarrow gg) = \frac{G_F\alpha^2_sm^3_h}{36\sqrt{2}\pi^3}
|\sum_{Q}A^h_Q(\tau_Q)|^2,
\end{equation}
\begin{equation}
A^h_Q(\tau) = \frac{3}{2}\tau [1 + (1 -\tau)f(\tau),
\end{equation}
\begin{equation}
argsin^2(\frac{1}{\sqrt{\tau}}), \,\,  \tau \geq 1,
\end{equation}
\begin{equation}
f(\tau) = -\frac{1}{4}[\log(\frac{1 +\sqrt{1 -\tau}}{1 -\sqrt{1 - \tau}} 
- i\pi]^2, \,\,  \tau < 1
\end{equation}
The parameter $\tau_Q = \frac{4m^2_Q}{m^2_h}$ is defined by the pole mass 
$m_Q$ of the heavy quark  in the loop. For large quark mass 
$ A^h_Q(\tau_Q) \rightarrow 1$. An account of the QCD radiative corrections
 (see Fig.5) 
gives for $m^2_h \ll 4m^2_Q$  \cite{40}
\begin{equation}
\Gamma(h \rightarrow gg(q), q\bar{q}g) = \Gamma_{LO}[\alpha_s^{(N_F)}(m_h)]
[1 + (\frac{95}{4} - \frac{7}{6}N_F) \frac{\alpha_s^{(N_F)}(m_h)}{\pi}]
\end{equation}
with $N_F = 5 $ light quark flavours.  It appears that radiative 
corrections are very large : the 
decay width is shifted by about (60 -70) percent upwards in the most 
interesting  
mass region  100 GeV $ \leq m_h \leq 500$ GeV. Three loop QCD corrections 
have been calculated in the limit of a heavy top quark \cite{41}.
They are positive 
and increase the full next leading order expression by 10 percent. 
Using the low-energy theorems  it is possible to calculate easily 
the electroweak $O(G_FM^2_t)$ corrections to the leading order Higgs boson 
decay width into two gluons \cite{42}
\begin{equation}
\Gamma(h \rightarrow gg) = \Gamma_{LO}(h \rightarrow gg)
[1 + \frac{G_FM^2_t}{8\sqrt{2}\pi^2}]
\end{equation}
Numerically they are negligible.

\begin{figure}[hbt]
\begin{center}
\setlength{\unitlength}{1pt}
\begin{picture}(180,100)(0,0)

\Gluon(100,20)(150,20){-3}{5}
\Gluon(100,80)(150,80){3}{5}
\ArrowLine(100,20)(100,80)
\ArrowLine(100,80)(50,50)
\ArrowLine(50,50)(100,20)
\DashLine(0,50)(50,50){5}
\put(-15,46){$h$}
\put(105,46){$t,b$}
\put(155,18){$g$}
\put(155,78){$g$}

\end{picture}  \\
\setlength{\unitlength}{1pt}
\caption[ ]{\label{fg:hgglodia} \it Diagrams contributing to $h\to gg$ at
lowest order.}
\end{center}
\end{figure}
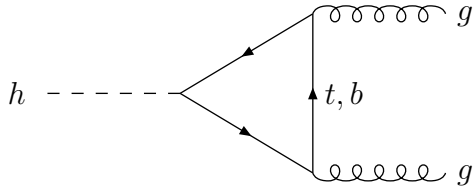

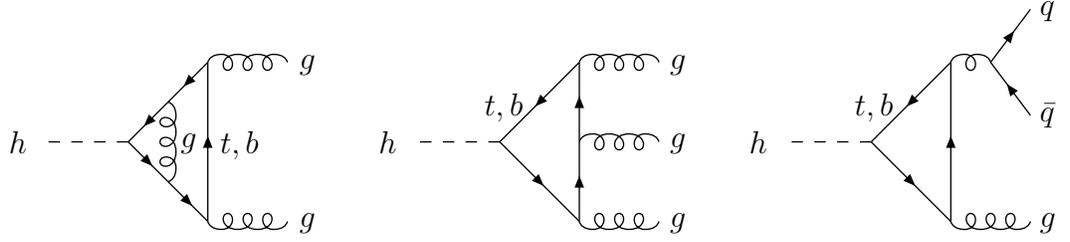
\begin{figure}[hbt]
\begin{center}
\setlength{\unitlength}{1pt}
\begin{picture}(400,100)(-20,0)

\Gluon(60,20)(90,20){-3}{3}
\Gluon(60,80)(90,80){3}{3}
\Gluon(45,35)(45,65){-3}{3}
\ArrowLine(60,20)(60,80)
\ArrowLine(60,80)(45,65)
\ArrowLine(45,65)(30,50)
\ArrowLine(30,50)(45,35)
\ArrowLine(45,35)(60,20)
\DashLine(0,50)(30,50){5}
\put(-15,46){$h$}
\put(65,46){$t,b$}
\put(95,18){$g$}
\put(95,78){$g$}
\put(50,48){$g$}

\Gluon(200,20)(230,20){-3}{3}
\Gluon(200,80)(230,80){3}{3}
\Gluon(200,50)(230,50){3}{3}
\ArrowLine(200,20)(200,50)
\ArrowLine(200,50)(200,80)
\ArrowLine(200,80)(170,50)
\ArrowLine(170,50)(200,20)
\DashLine(140,50)(170,50){5}
\put(125,46){$h$}
\put(165,60){$t,b$}
\put(235,18){$g$}
\put(235,48){$g$}
\put(235,78){$g$}

\Gluon(340,20)(370,20){-3}{3}
\Gluon(340,80)(355,80){3}{1}
\ArrowLine(355,80)(370,100)
\ArrowLine(370,60)(355,80)
\ArrowLine(340,20)(340,80)
\ArrowLine(340,80)(310,50)
\ArrowLine(310,50)(340,20)
\DashLine(280,50)(310,50){5}
\put(265,46){$h$}
\put(305,60){$t,b$}
\put(375,18){$g$}
\put(375,58){$\bar q$}
\put(375,98){$q$}

\end{picture}  \\
\setlength{\unitlength}{1pt}
\caption[ ]{\label{fg:hggdia} \it Typical diagrams contributing to the QCD
corrections to $h\to gg$.}
\end{center}
\end{figure}

Also very important is  the one-loop induced Higgs boson coupling
to two photons due to $W$- and $t$-quark exchanges in the loop (see Fig.6). 
The partial decay 
width can be written in the form
\begin{equation}
\Gamma(h \rightarrow \gamma\gamma) = \frac{G_F\alpha^2m^3_h}{128\sqrt{2}\pi^3}
|\sum_{f}N_{cf}e^2_fA^h_f(\tau_f) +A^h_W(\tau_W)|^2,
\end{equation}
where
\begin{equation}
A^h_f(\tau) = 2\tau[1 +(1-\tau)f(\tau)],
\end{equation}
\begin{equation}
A^h_W(\tau) = -[2 +3\tau +3\tau(2 - \tau)f(\tau)],
\end{equation}
$\tau_i = \frac{4 M^2_i}{m^2_h}$, $i = f, W$ 
and the function $f(\tau)$ is determined by the formulae
 (62,63). The $W$ loop gives 
the dominant contribution in the intermediate Higgs boson mass range. 
Two-loop QCD corrections to the quark loops have been calculated in 
refs.\cite{42}. QCD corrections rescale the lowest order  by a factor that 
depends on the ratio of the Higgs boson and quark masses 
\begin{equation}
A^h_Q(\tau_Q) \rightarrow A^h_Q(\tau_Q) \times [ 1 +  C_h(\tau_Q) 
\frac{\alpha_s}{\pi}]
\end{equation}
with $ C_h(\tau_Q) \rightarrow -1$ for $m^2_h \ll 4m^2_Q$.  QCD corrections to 
the two photon Higgs boson decay width numerically 
are not very big, of $O(10)~\%$.
Electroweak corrections are less than $1~\%$ \cite{43}.

\begin{figure}[hbt]
\begin{center}
\setlength{\unitlength}{1pt}
\begin{picture}(400,100)(0,0)

\Photon(60,20)(90,20){-3}{4}
\Photon(60,80)(90,80){3}{4}
\ArrowLine(60,20)(60,80)
\ArrowLine(60,80)(30,50)
\ArrowLine(30,50)(60,20)
\DashLine(0,50)(30,50){5}
\put(-15,46){$h$}
\put(70,46){$f$}
\put(95,18){$\gamma$}
\put(95,78){$\gamma$}

\Photon(210,20)(240,20){-3}{4}
\Photon(210,80)(240,80){3}{4}
\Photon(210,80)(210,20){3}{7}
\Photon(210,20)(180,50){3}{5}
\Photon(180,50)(210,80){3}{5}
\DashLine(150,50)(180,50){5}
\put(135,46){$h$}
\put(220,46){$W$}
\put(245,18){$\gamma$}
\put(245,78){$\gamma$}

\DashLine(300,50)(330,50){5}
\PhotonArc(345,50)(15,0,180){3}{6}
\PhotonArc(345,50)(15,180,360){3}{6}
\Photon(360,50)(390,80){3}{5}
\Photon(360,50)(390,20){3}{5}
\put(285,46){$h$}
\put(335,70){$W$}
\put(395,18){$\gamma$}
\put(395,78){$\gamma$}

\end{picture}  \\
\setlength{\unitlength}{1pt}
\caption[ ]{\label{fg:hgagalodia} \it Diagrams contributing to $h\to
\gamma \gamma$ at lowest order.}
\end{center}
\end{figure}
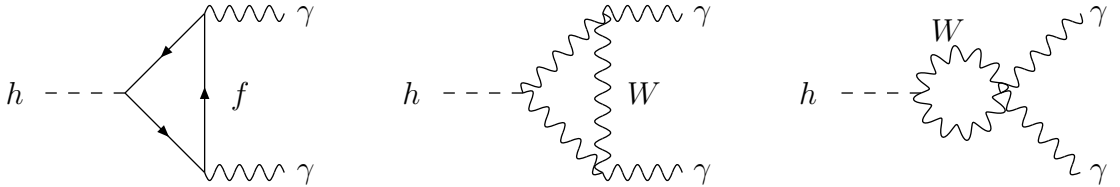

\begin{figure}[hbtp]

\vspace*{0.5cm}
\hspace*{1.0cm}
\begin{turn}{-90}%
\epsfxsize=8.5cm \epsfbox{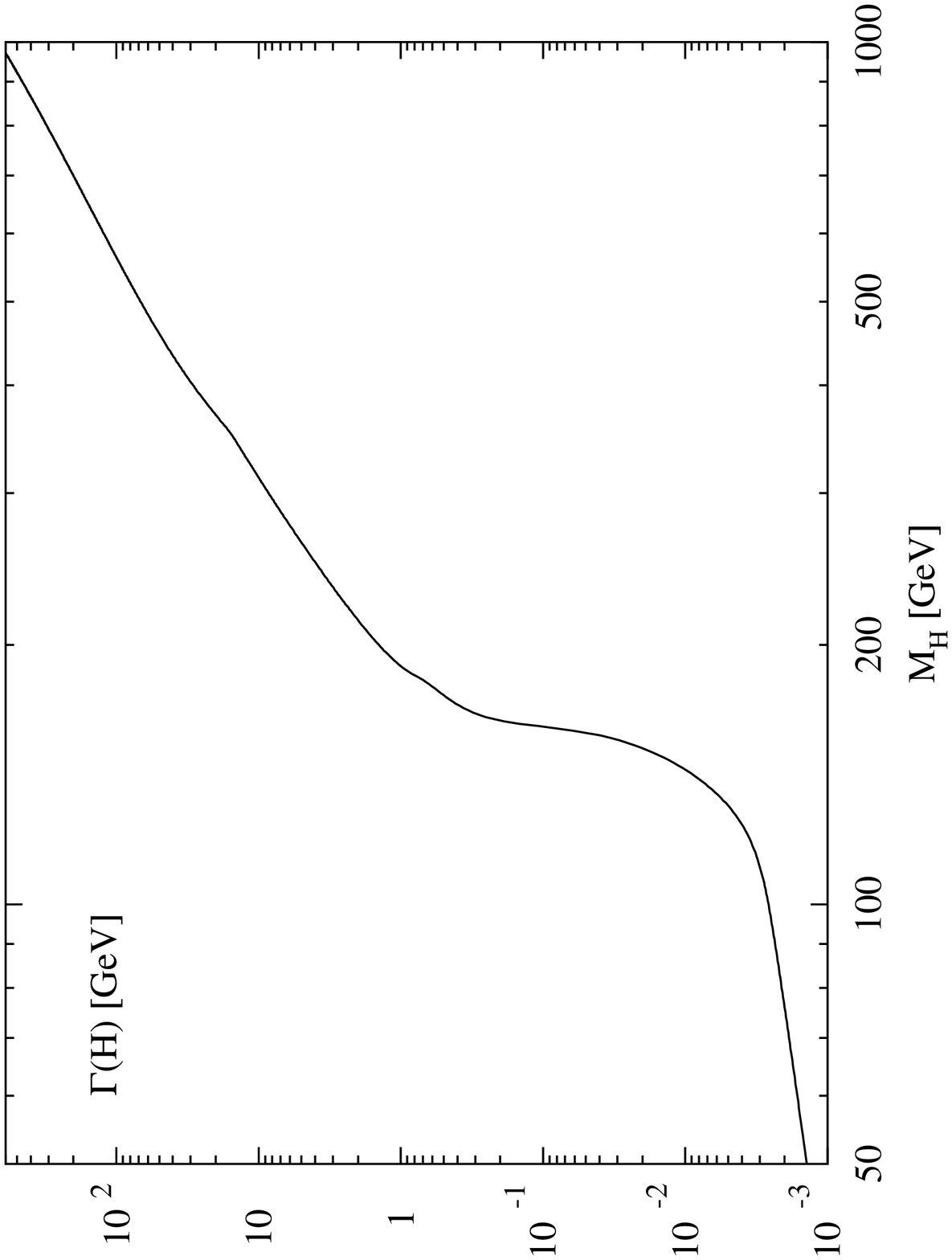}
\end{turn}

\vspace*{0.5cm}
\hspace*{1.0cm}
\begin{turn}{-90}%
\epsfxsize=8.5cm \epsfbox{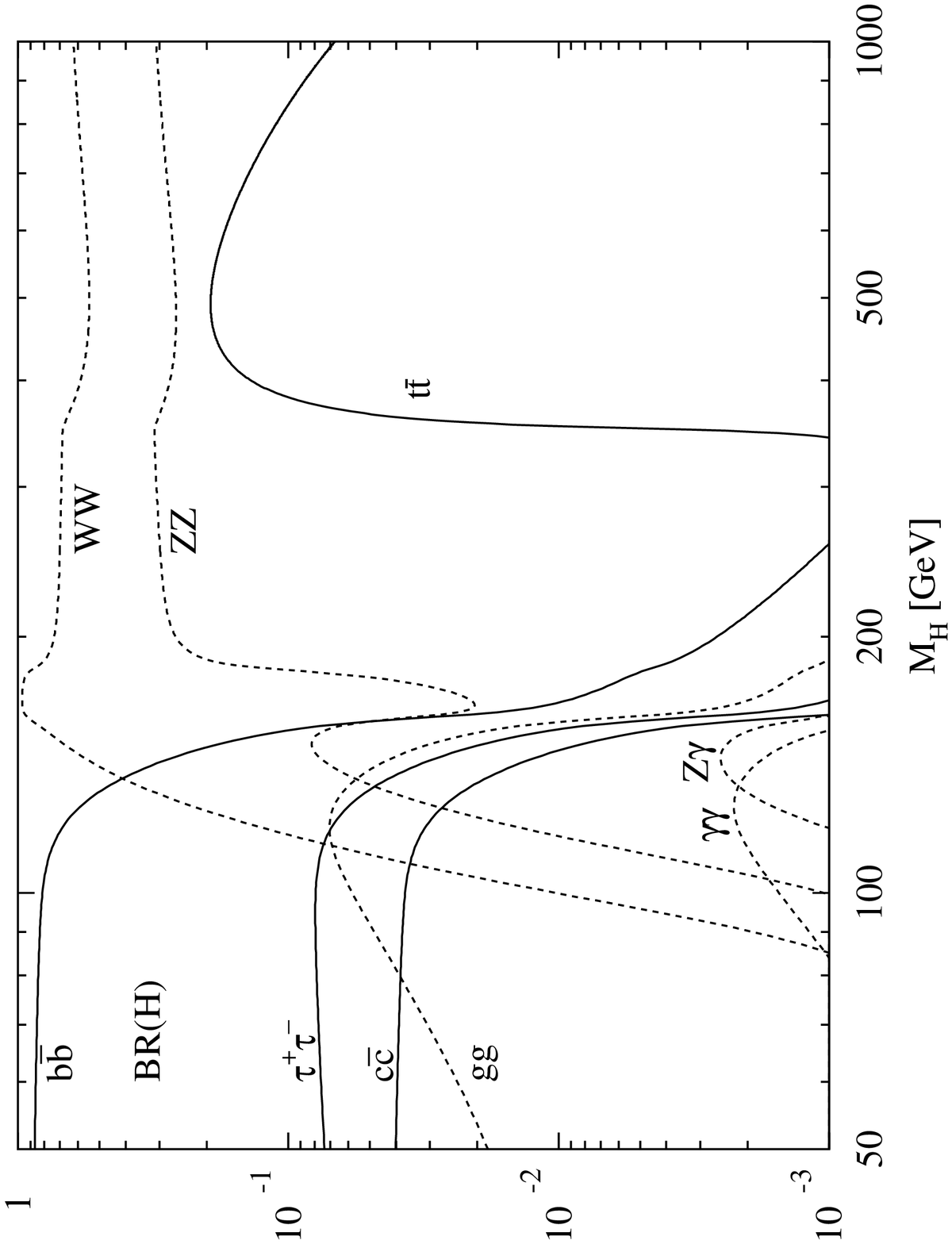}
\end{turn}
\vspace*{-0.0cm}

\caption[]{\label{fg:wtotbr} \it (a) Total decay width (in GeV) of the SM
Higgs boson as a function of its mass. (b) Branching ratios of the
dominant decay modes of the SM Higgs particle. All relevant higher-order
corrections are taken into account (ref.\cite{16}) ($H \equiv h$).}
\end{figure}

\section{Higgs boson search at LEP}

The process that was used for the direct search for the Higgs boson at LEP1
was the Bjorken process \cite{45}
\begin{equation}
e^+e^- \rightarrow Z \rightarrow (Z^* \rightarrow f\bar {f})h
\end{equation}
The differential decay width for the $Z \rightarrow (Z^* \rightarrow f\bar{f})
h$ reaction  normalized to $Z \rightarrow f \bar{f}$ decay is given by 
\cite{46}
\begin{equation}
\frac{\Gamma(Z  \rightarrow (Z^{*} \rightarrow f\bar{f})h) }
{\Gamma(Z \rightarrow f\bar{f})} = 
\frac{\alpha}{4\pi\sin^2\theta_W \cos^2\theta_W}
\frac{(1 -x +\frac{x^2}{12} +\frac{2r^2}{3})(x^2 - 4r^2)^{1/2}}
{(x - r^2)^2 + (\Gamma_Z/M_Z)^2}\,,
\end{equation}
where $x = 2E_h/M_Z$ and $r = m_h/M_Z$, the kinematical limits being 
$2r \leq x \leq 1 - r^2$. The energy of the Higgs boson $E_h$ is related 
to the invariant mass of the fermion pair $M_{f\bar{f}}$ (i.e. the invariant 
mass of the virtual $Z^*$ boson)
\begin{equation}
E_h = \frac{(M^2_Z + m^2_h - M^2_{f\bar{f}})}{2M_Z}
\end{equation} 
The Bjorken process with the decay 
of the virtual $Z$-boson  to $\mu^+\mu^-, e^+e^-, \nu \bar{\nu}$
 pairs is used for the Higgs boson search. The decay of the $Z^*$ to 
quark-antiquark pair is not useful due to large QCD background. The Higgs 
decay mode determines the Higgs signature in the detectors. Higgs bosons 
with low mass decay into $e^+e^-$ and $\mu^+\mu^-$ pairs, for intermediate 
mass they decay into light hadrons and $\tau^+\tau^-$ pairs, and for high 
mass they decay mainly into a $b\bar{b}$ quark-antiquark pair. The combined 
limit of the four LEP1 experiments(ALEPH, DELPHI, L3 and OPAL) on the 
Higgs boson mass is  \cite{47}
\begin{equation}
m_h \geq 65.4 ~GeV, \,\, 95\% ~C.L.
\end{equation}
      
At LEP2 with the total
energy $\sqrt{s} =130 -  200 $ GeV the dominant Higgs production
process \footnote{The $e^+e^- \rightarrow WW$ and $e^+e^- \rightarrow 
ZZ$ fusions are still negligible at LEP2 energies}
 is $e^+e^- \rightarrow hZ$ (``Higgs-Strahlung'' process). 
 The corresponding cross section
at tree level is given by \cite{48}
\begin{equation}
\sigma(e^+e^- \rightarrow hZ) =
\frac{\pi \alpha^{2}\lambda^{1/2}
(\lambda + 12sM^2_Z)
[1 + (1 - 4\sin^{2}{{\theta}_W})^2]}
{192s^2\sin^4{{\theta}_W} \cos^4{{\theta}_W}(s -M^2_Z)^2}\,,
\end{equation}
where $\lambda \equiv (s - m^2_h -M^2_Z)^2 - 4m^2_hM^2_Z$. One can
see that for a fixed value of $m_h$, the cross section is maximal
for $\sqrt{s} \approx m_Z +\sqrt{2}m_h$. 

There are important differences between the Higgs boson searches at LEP1 
and LEP2. The signal-to-background ratio is much better at LEP2. The large 
background rate at LEP1 required a very detailed simulation  of 
detector effects and rare background reactions. The dominant hadronic Higgs 
boson signature $(Z^*h \rightarrow q\bar{q}q\bar{q})$ was useless at LEP1 
due to large QCD background. 
While the expected  Higgs boson production at LEP1 involved a real  $Z$ 
decaying into a Higgs boson and a virtual $Z$ boson, at LEP2 the Higgs boson 
is produced in association with an on-shell $Z$ boson. This additional 
information about the final state $Z$ boson gives rise to better Higgs boson 
mass reconstruction and greater sensitivity for a Higgs boson signal due to 
better background rejection.        

Final state particles in the analysed 
Higgs boson channels at LEP2 are
\begin{equation}
e^+e^- \rightarrow (Z \rightarrow q\bar{q}, \nu\bar{\nu}, e^+e^-, 
\mu^+\mu^-,\tau^+\tau^-)(h \rightarrow b\bar{b}, \tau^+\tau^-)
\end{equation}    
Thus the three typical signatures are 

(a) two $b$-jets + a charged lepton pair ($Z \rightarrow \mu^+\mu^-
(e^+e^-)$, $h \rightarrow b\bar{b}$);

(b) two $b$-jets plus missing transverse energy ($Z \rightarrow \nu \bar{\nu}$ 
,$h \rightarrow b\bar{b})$;

(c) four jets with at least two $b$-jets or two $\tau$-jets
($Z \rightarrow q\bar{q}$, $h \rightarrow b\bar{b}$ or ($hZ \rightarrow 
q\bar{q}\tau^+\tau^-$)).

Standard model background to these signatures is well known 
and it is under control \cite{48,49}. 
For example, the Higgs boson production cross section at $\sqrt{s} =
189~GeV$  for $m_h = 95~GeV$ is $0.18~pb$, whereas the 
main background cross sections are $98~pb~(e^+e^- \rightarrow q\bar{q})$, 
$16~pb (e^+e^- \rightarrow WW)$, $0.62~pb (e^+e^- \rightarrow ZZ)$.
   
1998 LEP2 run with full energy $\sqrt s = 189~GeV$ and with 
$L \approx 170~pb^{-1}/exp$ allowed to deduce the 
following $95\, \%\,C.L.$ lower Higgs boson mass bounds 
\cite{50} - \cite{54}
$$ m_h > 90.2~GeV~(ALEPH),$$
$$ m_h > 95.2~GeV~(DELPHI),$$
$$ m_h > 95.3~GeV~(L3),$$
$$ m_h > 91.0~GeV~(OPAL)$$. 
Note that an additional account of 1999 data with 
integrated luminosities $29~fb^{-1}$ and $69.5~fb^{-1}$ at
$\sqrt{s} = 191.6~GeV$ and $\sqrt{s} = 195.6~GeV$ allowed the 
ALEPH Collaboration deduce the Higgs boson mass bound 
$m_h > 98.8~GeV$ \cite{55} Recent preliminary combined limit of 4 LEP2 
experiments with $\sqrt{s} \leq 195.6~GeV$ \cite{56} gives 
$m_h > 102.6~GeV$ at $95\, \% \, C.L.$. 

 LEP2 run with total 
energy $\sqrt{s} = 200~GeV$ and with total luminosity $L_t = 200~pb^{-1}$ 
for each experiment will be able  to discover standard Higgs boson with a mass 
up to 107 GeV \cite{57}.

\section{Higgs boson production at hadron 
supercolliders}

Typical processes that can be exploited to produce Higgs bosons in 
hadron supercolliders are:

gluon fusion:\,\, $gg \rightarrow h$

WW, ZZ fusion:\,\, $W^+W^-, ZZ \rightarrow h$ 

Higgs-strahlung off W, Z: \, \, $q\bar{q} W,Z \rightarrow W,Z + h$

Higgs bremsstrahlung off top: \, \, $q\bar{q}, gg \rightarrow t \bar{t} + h$

Gluon fusion plays a dominant role at the LHC throughout the 
entire Higgs boson mass range 
of the SM whereas the $WW/ZZ$ fusion process becomes increasingly 
important with Higgs boson rising. The last two reactions are important only 
for light Higgs boson masses. 
                                                                             
The gluon-fusion mechanism \cite{58} (see Fig.8)
\begin{equation}
pp \rightarrow gg \rightarrow h
\end{equation}
is the dominant production mechanism of the Higgs boson at the LHC for 
Higgs boson mass up to 1 TeV. The gluon coupling to the Higgs boson in the 
SM is mediated by triangular loops of top and bottom quarks. The 
corresponding form factor approaches a non-zero value for large 
loop-quark masses. At lowest order the partonic cross section can be 
expressed by the gluonic width of the Higgs boson
\begin{equation}
\hat{\sigma}_{LO}(gg \rightarrow h) = \sigma_0 m^2_h \delta(\hat{s} -m^2_h),
\end{equation}
\begin{equation}
\sigma_0 = \frac{\pi^2}{8m^2_h}\Gamma_{LO}(h \rightarrow gg)
= \frac{G_F\alpha^2_s}{288\sqrt{2}\pi}|\sum _{Q}A^h_Q(\tau_Q)|^2 ,
\end{equation}     
where $\tau_Q = \frac{4M^2_Q}{m^2_h}$, $\hat{s}$ denotes the partonic 
system of mass energy squared and the form factor $A^h_Q$ is determined 
by the formulae (62,63 ). In the narrow-width approximation hadronic 
cross section can be written in the form 
\begin{equation}
\sigma_{LO}(pp \rightarrow h +...) = \sigma_0 \tau_h \frac{dL^{gg}}{d\tau_h},
\end{equation}
where $\frac{dL^{gg}}{d\tau_h}$ denotes $gg$ luminosity of the pp collider 
with $\tau_h = \frac{m^2_h}{s}$.  The QCD corrections to the gluon fusion 
process (see Fig.9)are  essential \cite{59} . They stabilise the 
theoretical predictions 
for the cross section when the renormalization and factorisation scales are 
varied. Moreover, they are large and positive, thus increasing  
the production cross section for Higgs bosons. The QCD corrections consist of 
virtual corrections to the basic process $gg \rightarrow h$ and of real 
corrections due to reactions $gg \rightarrow hg$, $qq \rightarrow hq$ and 
$q\bar{q} \rightarrow hg$. The virtual corrections rescale the 
lowest-order fusion cross section with a coefficient that depends only on 
the ratios of the Higgs and quark masses. 
The next to leading order
for the hadronic cross section can be represented in the form \cite{59}
\begin{equation}
\sigma(pp \rightarrow h + ...) = \sigma_0 [1 +  C \frac{\alpha_s}{\pi}]
\tau_h \frac{dL^{gg}}{d\tau_h} + \Delta \sigma_{gg} + 
\Delta \sigma_{qg} + \Delta \sigma_{q\bar{q}}
\end{equation}
The calculation has been performed \cite{59} in the $\bar{MS}$ scheme. 
The mass $M_Q$ is identified with the pole quark mass and the 
renormalization scale in $\alpha_s$ and the factorisation scale of the 
parton densities is fixed at the Higgs boson mass. The coefficient 
$C(\tau_Q)$ denotes the finite part of the virtual two-loop corrections. 
The finite parts of the hard contributions from gluon radiation in 
$gg$ scattering , $gq$ scattering and $q\bar{q}$ annihilation are 
presented in the form \cite{59}
\begin{eqnarray}
&&\Delta \sigma_{gg} = \int ^1_{\tau_h} d\tau \frac{dL^{gg}}{d\tau}
\times \frac{\alpha_s}{\pi}\sigma_0 
[-zP_{gg}(z)\log z + d_{gg}(z, \tau_Q) + \\ \nonumber 
&&12[(\frac{\log(1-z)}{(1-z)}_{+} -z[2 - z(1-z)]\log(1-z)]]\,,
\end{eqnarray}
\begin{equation}
\Delta \sigma_{gq} = \int^1_{\tau_h}\sum_{q,\bar{q}}\frac{dL^{gq}}{d\tau} 
\times 
\frac{\alpha_s}{\pi}\sigma_0[-\frac{z}{2}P_{gq}(z)\log
\frac{z}{(1-z)^2} +d_{gq}(z, \tau_Q)]\,,
\end{equation}
\begin{equation}
\Delta \sigma_{q\bar{q}} = \int^1_{\tau_h}d\tau \sum_{q} \frac{dL^{q\bar{q}}}
{d\tau} \times \frac{\alpha_s}{\pi}\sigma_0d_{q\bar{q}}(z,\tau_Q)\,,
\end{equation}
where $z = \tau_h/\tau = m^2_h/\hat{s}$, $P_{gg} $ and $P_{gq}$ are 
Altarelli-Parisi splitting functions. In the heavy quark limit one can 
find that \cite{59}
\begin{equation}
C(\tau_Q) \rightarrow \pi^2 + 5.5 ,
\end{equation}
\begin{equation}
d_{gg}(z,\tau_Q) \rightarrow -5.5(1-z)^3 ,
\end{equation}
\begin{equation}
d_{gq}(z, \tau_Q) \rightarrow \frac{2}{3}z^2 - (1-z)^2 ,
\end{equation}
\begin{equation}
d_{q\bar{q}}(z, \tau_Q) \rightarrow \frac{32}{27}(1-z)^3
\end{equation}

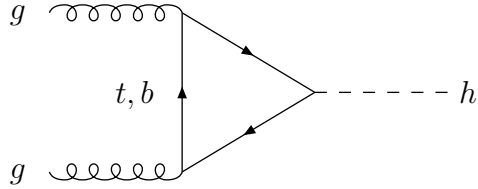
\begin{figure}[hbt]
\begin{center}
\setlength{\unitlength}{1pt}
\begin{picture}(180,90)(0,0)

\Gluon(0,20)(50,20){-3}{5}
\Gluon(0,80)(50,80){3}{5}
\ArrowLine(50,20)(50,80)
\ArrowLine(50,80)(100,50)
\ArrowLine(100,50)(50,20)
\DashLine(100,50)(150,50){5}
\put(155,46){$h$}
\put(25,46){$t,b$}
\put(-15,18){$g$}
\put(-15,78){$g$}

\end{picture}  \\
\setlength{\unitlength}{1pt}
\caption[ ]{\label{fg:gghlodia} \it Diagram contributing to the
formation of Higgs bosons in gluon-gluon collisions
at lowest order.}
\end{center}
\end{figure}

\begin{figure}[hbt]
\begin{center}
\setlength{\unitlength}{1pt}
\begin{picture}(450,100)(-10,0)

\Gluon(0,20)(30,20){3}{3}
\Gluon(0,80)(30,80){3}{3}
\Gluon(30,20)(60,20){3}{3}
\Gluon(30,80)(60,80){3}{3}
\Gluon(30,20)(30,80){3}{5}
\ArrowLine(60,20)(60,80)
\ArrowLine(60,80)(90,50)
\ArrowLine(90,50)(60,20)
\DashLine(90,50)(120,50){5}
\put(125,46){$h$}
\put(65,46){$t,b$}
\put(-10,18){$g$}
\put(-10,78){$g$}
\put(15,48){$g$}

\Gluon(160,100)(210,100){3}{5}
\Gluon(210,100)(270,100){3}{6}
\Gluon(160,0)(210,0){3}{5}
\Gluon(210,100)(210,60){3}{4}
\ArrowLine(210,0)(210,60)
\ArrowLine(210,60)(240,30)
\ArrowLine(240,30)(210,0)
\DashLine(240,30)(270,30){5}
\put(275,26){$h$}
\put(215,26){$t,b$}
\put(150,-2){$g$}
\put(150,98){$g$}
\put(195,78){$g$}
\put(275,98){$g$}

\ArrowLine(310,100)(360,100)
\ArrowLine(360,100)(420,100)
\Gluon(310,0)(360,0){3}{5}
\Gluon(360,100)(360,60){3}{4}
\ArrowLine(360,0)(360,60)
\ArrowLine(360,60)(390,30)
\ArrowLine(390,30)(360,0)
\DashLine(390,30)(420,30){5}
\put(425,26){$h$}
\put(365,26){$t,b$}
\put(300,-2){$g$}
\put(300,98){$q$}
\put(425,98){$q$}
\put(345,78){$g$}

\end{picture}  \\
\setlength{\unitlength}{1pt}
\caption[ ]{\label{fg:gghqcddia} \it Typical diagrams contributing to the
virtual/real QCD corrections to $gg\to h$.}
\end{center}
\end{figure}
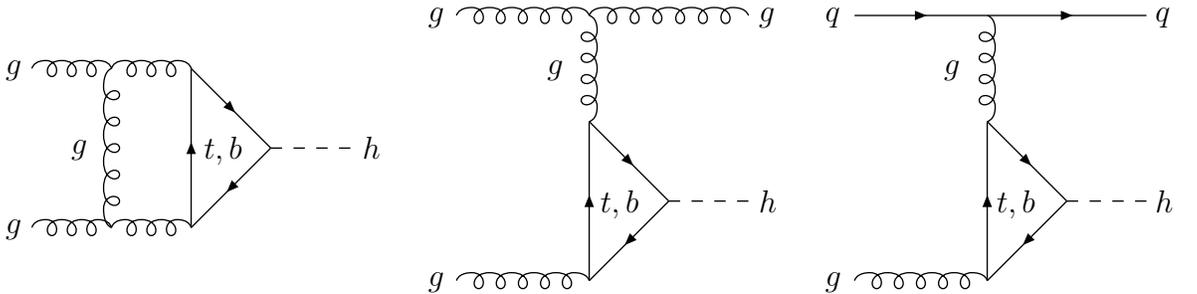

The size of the radiative corrections can be parametrised by defining the $K$ 
factor as $K = \sigma_{NLO}/\sigma_{LO}$. The results of the calculations 
are presented in Fig. 10. The virtual and the real corrections 
for the $gg$ collisions are the most important, they are large and positive. 
After including higher order QCD corrections the dependence of the 
cross section on the renormalization and factorisation scales is reduced 
from the level of $O(1)$ to a level of about $O(0.2)$.

The theoretical prediction for the Higgs boson production cross section 
is presented in Fig.11 for the LHC as a function of the Higgs boson mass. 
The cross section decreases with increasing of the Higgs boson mass  
mainly due to the decrease 
of $gg$ partonic luminosity for large invariant masses.

\begin{figure}[hbt]

\vspace*{0.4cm}
\hspace*{2.0cm}
\begin{turn}{-90}%
\epsfxsize=7cm \epsfbox{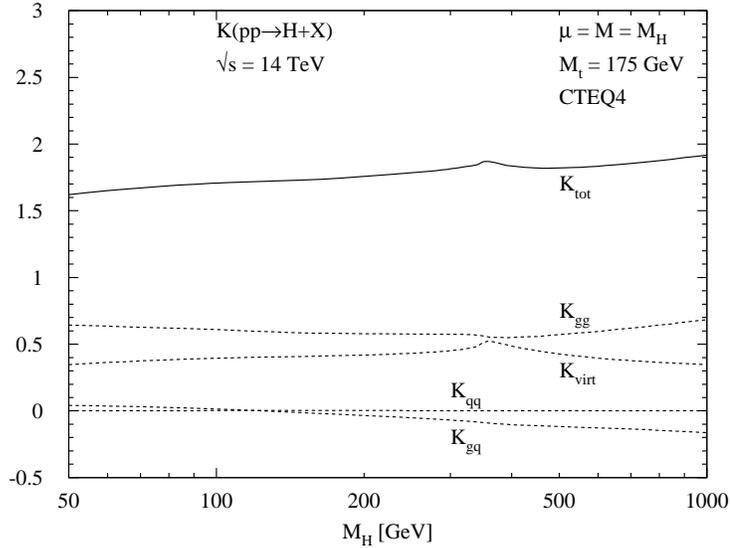}
\end{turn}
\vspace*{-0.2cm}

\caption[]{\label{fg:gghk} \it K factors of the QCD-corrected gluon-fusion
cross section $\sigma(pp \to h+X)$ at the LHC with c.m.~energy $\sqrt{s}=14$
TeV.  The renormalization and
factorisation scales have been identified with the Higgs mass,  
and  CTEQ4 parton densities have been adopted (ref.\cite{16}) ($H \equiv h$) .}
\end{figure}

\begin{figure}[hbt]

\vspace*{0.5cm}
\hspace*{2.0cm}
\begin{turn}{-90}%
\epsfxsize=7cm \epsfbox{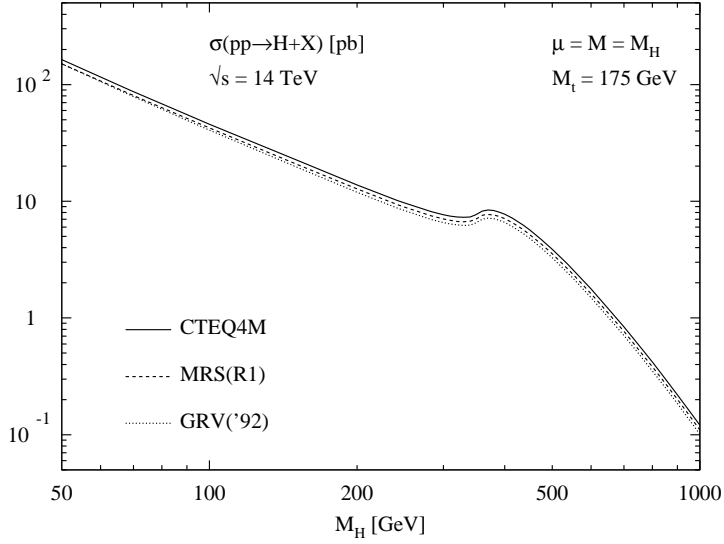}
\end{turn}
\vspace*{0.0cm}

\caption[]{\label{fg:gghparton} \it The cross 
section for the production of Higgs bosons;  three
different sets of parton densities are shown
[CTEQ4M, MRS(R1) and GRV('92)] (ref.\cite{16}) ($H \equiv h$).}
\end{figure}

The second important process for the Higgs boson production at the LHC 
is vector-boson fusion (see Fig.12),  
$W^+W^-(ZZ) \rightarrow h$ \cite{60}. For large 
Higgs boson mass this mechanism becomes competitive to 
gluon fusion; for intermediate masses the cross section is smaller 
by about an order of magnitude. For large 
Higgs boson mass the $W$ and $Z$ bosons are predominantly longitudinally 
polarised.  
At high energies, the equivalent particle spectra of the longitudinal 
$W$, $Z$ bosons in quark beam have the form \cite{16}
\begin{equation}
f^W_L(x) = \frac{G_FM^2_W}{2\sqrt{2}\pi^2}\frac{1-x}{x}\,,
\end{equation}
\begin{equation}
f^Z_L(x) = \frac{G_FM^2_Z}{2\sqrt{2}\pi^2}[(I^q_3 -2e_q\sin^2\theta_W)^2 +
(I^q_3)^2]\frac{1-x}{x}\,,
\end{equation}
where $x$ is the fraction of energy transferred  from the quark to the 
$W$, $Z$ boson in the splitting process $q \rightarrow q + W/Z$. The 
$WW$ and $ZZ$ luminosities are presented in the form:
\begin{equation}
\frac{dL^{WW}}{d\tau_W} = \frac{G^2_FM^4_W}{8\pi^4}[ 2 - \frac{2}{\tau_W} 
- \frac{1 +\tau_W}{\tau_W}\log \tau_W]\,,
\end{equation}

\begin{eqnarray}
&&\frac{dL^{ZZ}}{d\tau_Z} = \frac{G^2_FM^4_Z}{8\pi^4}[(I^q_3 -2e_q\sin^2
\theta_W)^2 +(I^q_3)^2][(I^{q^{'}})^2 - 2e _{q^{`}}\sin^2\theta_W)^2 
+ (I^{q^{`}}_3)^2] \cdot  \\ \nonumber 
&&[2 - \frac{2}{\tau_Z} - \frac{1 + \tau_Z}{\tau_Z}
\log\tau_Z]\,,
\end{eqnarray}

where  $ \tau_V = \frac{M^2_{VV}}{s}$. Denoting the parton cross section for 
$WW, ZZ \rightarrow h$ by $\hat{\sigma}_0$ with
\begin{equation}
\hat{\sigma}_0(VV \rightarrow h) =  \sigma_0 \delta(1 -m^2_h/\hat{s}),
\end{equation}
\begin{equation}
\sigma_0 = \sqrt{2}\pi G_F\,,
\end{equation}
the cross sections for the Higgs boson production  in quark-quark  and 
hadron-hadron collisions are presented in the form \cite{16} 
\begin{equation}
\hat{\sigma}(qq \rightarrow qqh) = \frac{dL^{VV}}{d\tau_V}\sigma_0\,,
\end{equation}
\begin{equation}
\sigma(qq^{'} \rightarrow VV \rightarrow h) 
= \int^1_{m^2_h/s}\, d\tau \sum_{q,q^{'}}\frac{dL^{qq^{'}}}{d\tau}\hat{\sigma}
(qq^{'} \rightarrow qq^{'}h; \hat{s} =\tau s)\,.
\end{equation}

\begin{figure}[hbt]
\begin{center}
\setlength{\unitlength}{1pt}
\begin{picture}(120,110)(0,0)

\ArrowLine(0,0)(50,0)
\ArrowLine(50,0)(100,0)
\ArrowLine(0,100)(50,100)
\ArrowLine(50,100)(100,100)
\Photon(50,0)(50,50){3}{5}
\Photon(50,50)(50,100){3}{5}
\DashLine(50,50)(100,50){5}
\put(105,46){$h$}
\put(-15,-2){$q$}
\put(-15,98){$q$}
\put(55,21){$W,Z$}
\put(55,71){$W,Z$}

\end{picture}  \\
\setlength{\unitlength}{1pt}
\caption[ ]{\label{fg:vvhlodia} \it Diagram contributing to $qq \to qqV^*V^*
\to qqh$ at lowest order.}
\end{center}
\end{figure}
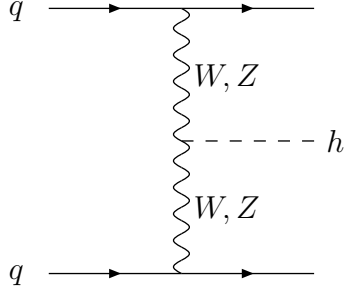

Higgs-strahlung $q\bar{q} \rightarrow V^* \rightarrow Vh$ $(V = W,Z)$ 
(see Fig.13) is a 
very important process for the search of light Higgs boson at the 
TEVATRON and LHC. Though the cross section is smaller than for gluon fusion, 
leptonic decays of electroweak vector bosons are extremely useful to filter 
Higgs boson signal from huge background. The corresponding formulae 
for the cross section are contained in \cite{61}.

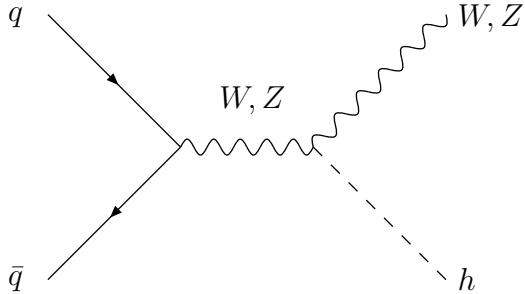
\begin{figure}[hbt]
\begin{center}
\setlength{\unitlength}{1pt}
\begin{picture}(160,120)(0,-10)

\ArrowLine(0,100)(50,50)
\ArrowLine(50,50)(0,0)
\Photon(50,50)(100,50){3}{5}
\Photon(100,50)(150,100){3}{6}
\DashLine(100,50)(150,0){5}
\put(155,-4){$h$}
\put(-15,-2){$\bar q$}
\put(-15,98){$q$}
\put(65,65){$W,Z$}
\put(155,96){$W,Z$}

\end{picture}  \\
\setlength{\unitlength}{1pt}
\caption[ ]{\label{fg:vhvlodia} \it Diagram contributing to $q\bar q \to V^*
\to Vh$ at lowest order.}
\end{center}
\end{figure}

The process $gg, q\bar{q} \rightarrow t \bar{t} h$ (see Fig.14) 
is relevant for small 
Higgs boson masses. The analytical expression for the parton cross section is 
quite involved \cite{62}. Note that Higgs boson bremsstrahlung off top 
quarks is an interesting process for measurements  
of the fundamental $h\bar{t}t$ Yukawa coupling. The cross section $\sigma(pp 
\rightarrow t\bar{t}h)$ is directly proportional to the square of this 
coupling constant.

\begin{figure}[hbt]
\begin{center}
\setlength{\unitlength}{1pt}
\begin{picture}(360,120)(0,-10)

\ArrowLine(0,100)(50,50)
\ArrowLine(50,50)(0,0)
\Gluon(50,50)(100,50){3}{5}
\ArrowLine(100,50)(125,75)
\ArrowLine(125,75)(150,100)
\ArrowLine(150,0)(100,50)
\DashLine(125,75)(150,50){5}
\put(155,46){$h$}
\put(-15,98){$q$}
\put(-15,-2){$\bar q$}
\put(65,65){$g$}
\put(155,98){$t$}
\put(155,-2){$\bar t$}

\Gluon(250,0)(300,0){3}{5}
\Gluon(250,100)(300,100){3}{5}
\ArrowLine(350,0)(300,0)
\ArrowLine(300,0)(300,50)
\ArrowLine(300,50)(300,100)
\ArrowLine(300,100)(350,100)
\DashLine(300,50)(350,50){5}
\put(355,46){$h$}
\put(235,98){$g$}
\put(235,-2){$g$}
\put(355,98){$t$}
\put(355,-2){$\bar t$}

\end{picture}  \\
\setlength{\unitlength}{1pt}
\caption[ ]{\label{fg:httlodia} \it Typical diagrams contributing to
$q\bar q/gg \to ht\bar t$ at lowest order.}
\end{center}
\end{figure}
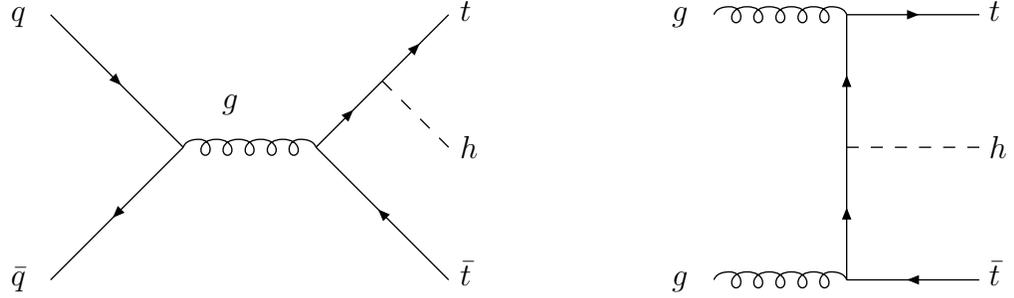

One can say that three classes of processes can be distinguished. The 
gluon fusion of Higgs boson is a universal process, dominant over the 
entire Higgs boson mass range. Higgs-strahlung of electroweak $W,Z$ bosons 
or top quarks is important for light Higgs boson. The $WW/ZZ$ fusion channel, 
by contrast, becomes rather important in the upper part of the Higgs boson 
mass. An overview of the production cross section for the Higgs boson at the 
LHC is presented in Fig. 15.

\begin{figure}[hbt]

\vspace*{0.5cm}
\hspace*{0.0cm}
\begin{turn}{-90}%
\epsfxsize=10cm \epsfbox{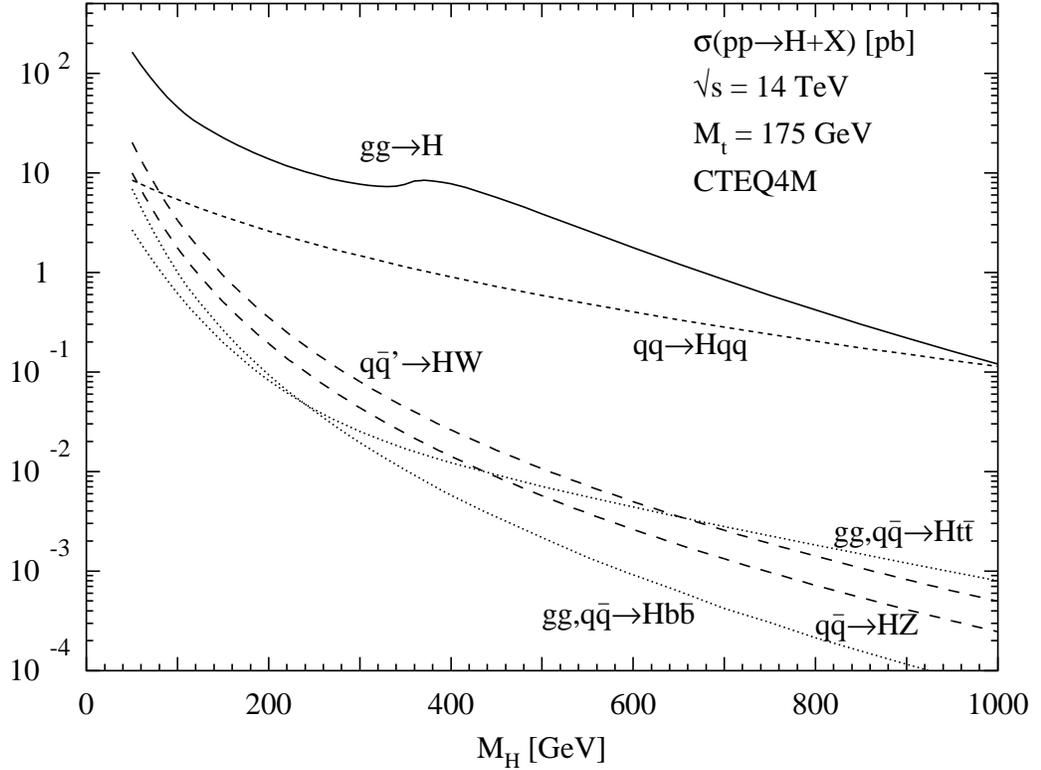}
\end{turn}
\vspace*{0.0cm}

\caption[]{\label{fg:lhcpro} \it Higgs production cross sections at the LHC
 for the various production mechanisms as a function of the
Higgs mass. The full QCD-corrected results for the gluon fusion $gg
\to h$, vector-boson fusion $qq\to VVqq \to hqq$, vector-boson bremsstrahlung
$q\bar q \to V^* \to hV$ and associated production $gg,q\bar q \to ht\bar t,
hb\bar b$ are shown \cite{16} ($H \equiv h$).}
\end{figure}

\section{Search for the Higgs boson at Tevatron}

It is expected \cite{63,64} that upgrated Fermilab Tevatron (TEV22) will 
start in 
2000 year with the full energy $\sqrt s = 2 ~TeV$ and the full luminosity 
for each experiment during 3 years of exploitation will be $L_t = 
2~fb^{-1}$. There are also plans to increase luminosity to have 
$L_t = 30~fb^{-1}$ (TEV33) by 2006. 
The most interesting process for the search for standard Higgs boson at 
the Tevatron is Higgs-strahlung off  W, Z bosons 
$q\bar{q} \rightarrow W^*/Z^* \rightarrow W/Z + h $. For the 
Higgs boson mass $ 100 ~GeV \leq m_h \leq 140 ~GeV$ the cross section is 
between 0.5 pb and 0.1 pb. The QCD corrections 
for Higgs-strahlung coincide with those of the  Drell-Yan process 
and increase tree level cross section approximately by 30 percent. 
The most promising signatures are
\begin{equation}
p\bar{p} \rightarrow (h \rightarrow b\bar{b})(W \rightarrow l  \nu, jets) 
+ anything,
\end{equation}
\begin{equation}
p \bar{p} \rightarrow (h \rightarrow b \bar{b})( Z \rightarrow l^+l^-, 
\nu \bar{\nu}) + anything.
\end{equation} 
The $b \bar{b}$ decay of the Higgs boson adds powerful background rejection 
based on $b$-tagging especially at low Higgs boson mass, below $\sim 130~GeV$ 
where that decay dominates.
Other very promising signature \cite{65} is the use of $h \rightarrow 
W^{*}W^{*} \rightarrow l\bar{\nu}\bar{l}\nu $ 
decay mode with the dominant 
gluon-gluon Higgs boson fusion production mechanism. The main conclusion of 
the ref. \cite{65} is that for an integrated luminosity 
of $30~fb^{-1}$ the Higgs boson signal should be observable at a 
$3\sigma$ level or better for the mass range $145~GeV \leq m_h \leq 180~GeV$ 
and for $95 \, \% $ percent confidence level exclusion, the mass reach is 
$135~GeV \leq m_h \leq 190~GeV$. 
One can say that     
at TEV33 run with the full luminosity $L_t = 30 fb^{-1}$ it would be possible 
to discover the  Higgs boson at $ \geq 3\sigma$ level at least 
with a mass up to $(180-190)~ GeV$ \cite{63}-\cite{65}.  
The Higgs boson discovery potential of Tevatron Collider is shown in 
Fig.16.

\begin{figure}[hbt]

\vspace*{0.5cm}
\hspace*{0.0cm}
\epsfxsize=10cm \epsfbox{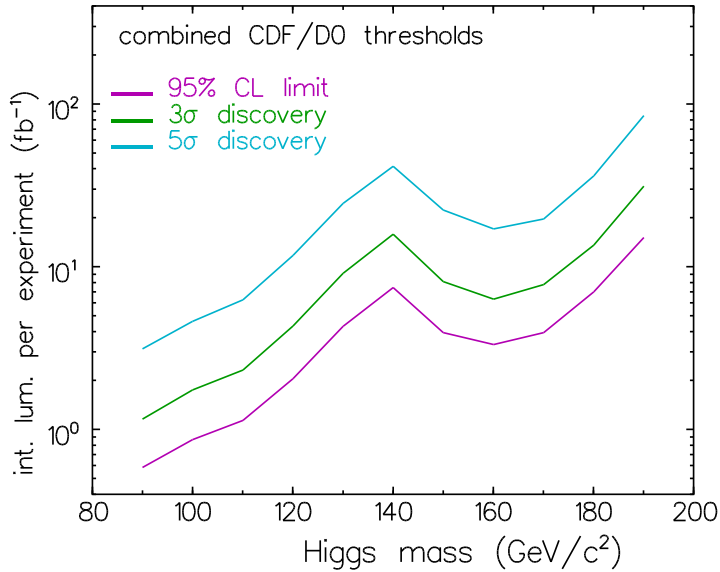}
\vspace*{0.0cm}

\caption[]{\label{fg:higgssensitivity} \it Luminosity required as a 
function of Higgs mass to achieve different levels of sensitivity to the 
standard-model Higgs boson.From the upper curve corresponds to a $5~\sigma$ 
discovery, the middle a $3~\sigma$ signal and the lower a 95\% exclusion 
limit. These limits require two experiments, Bayesian statistics are used 
to combine the channels and   include the improved sensitivity which would 
come from multivariate analysis techniques (ref.\cite{64}).}
\end{figure}

\section{LHC detectors}

The LHC(Large Hadron Collider) \cite{66}-\cite{69} which
will be the biggest particle accelerator complex ever built in the
World  will accelerate two proton beams with the total energy $\sqrt{s} =
14~TeV$. At low luminosity stage (first two-three years of operation)
the luminosity is planned to be $L_{low} = 10^{33}~cm^{-2}s^{-1}$
with total luminosity $L_{tot} = 10^{4}~pb^{-1}$ per year.
At high luminosity stage the luminosity is planned to be
$L_{high} = 10^{34}~cm^{-2}s^{-1}$ with total luminosity
$L_{tot} = 10^{5}~pb^{-1}$ per year. The LHC will start to
work in 2005 year. There are planned to be two big general purpose 
detectors at LHC CMS(Compact Muon Solenoid) and 
ATLAS(A Toroidal LHC Apparatus).  

The scientific program at the LHC consists in many goals \cite{66} - 
\cite{69}.
One of the most important tasks for the LHC is the quest for
the origin of the spontaneous symmetry-breaking mechanism
in the electroweak sector of the (SM).
As it has been mentioned before, all the renormalizable models of electroweak
interactions are based on the use of the gauge symmetry
breaking. As a consequence of the electroweak symmetry breaking and
the renormalizability of the theory there must be neutral
scalar particle(Higgs boson) in the
spectrum. So the discovery of the Higgs boson will be the
check of the spontaneous symmetry breaking and the renormalizability
of the theory and therefore there are no doubts that the Higgs boson 
discovery  is the supergoal 
number 1 for the LHC.  The Higgs boson 
search is therefore used as a first benchmark
for the detector optimisation for both CMS and ATLAS. For the SM
Higgs boson, the detector has to be sensitive to the
following processes in order to cover the full mass range
above the expected LEP2 discovery limit of $(105 - 110)~GeV$:

A. $h \rightarrow \gamma \gamma$ mass range  $90~ GeV \le m_{h} \leq 
150~GeV$.

B. $h \rightarrow b\bar{b}$ from $Wh, Zh, t\bar{t}h$ using
$l^{\pm}(l^{\pm} = e^{\pm}$ or $\mu^{\pm})$- tag and
b-tagging in the mass range $80~ GeV \leq  m_{h} \le 100 ~GeV$.

C. $h \rightarrow ZZ^{*} \rightarrow 4l^{\pm}$ for mass range
$130 ~GeV \le m_{h} \le 2m_{Z}$.

D. $h \rightarrow ZZ \rightarrow 4l^{\pm}, 2l^{\pm}2\nu$ for the
mass range $m_{h} \ge 2m_{Z}$.

E. $h \rightarrow WW, ZZ \rightarrow l^{\pm}\nu$ 2 jets, $2l^{\pm}$
2 jets, using tagging of forward jets for $m_{h}$ up to $\; 1$ TeV.

In minimal supersymmetric extension of the standard model(MSSM)
there is a family of Higgs particles $(H^{\pm}, h, H$ and $A$).
So in addition to  the standard Higgs boson signatures the
MSSM Higgs searches are based on the following processes:

F. $ A \rightarrow \tau^{+}\tau^{-} \rightarrow e \mu$ plus $\nu ' s$,
or $A \rightarrow \tau^{+} \tau^{-} \rightarrow l^{\pm}$
plus hadrons plus $\nu 's$.

G. $H^{\pm} \rightarrow \tau^{\pm} \nu$ from $t\bar{t} \rightarrow
H^{\pm}W^{\mp}b\bar{b}$ and $H^{\pm} \rightarrow 2$ jets, using a
$l^{\pm}$- tag and $b$-tagging.

The observable cross sections for most of those processes are small
$(1 - 100) ~pb$ over a large part of the mass range. So it is
necessary to work at high luminosity and to maximise
the detectable rates above backgrounds by high-resolution
measurements of electrons, muons and photons.

For the $H^{\pm}$ and $A$ signatures in the case of the MSSM,
high  performance detector capabilities are required in addition for
the measurements which are expected to be best achieved at initial
luminosities with a low level of overlapping events, namely secondary
vertex detection for $\tau$-leptons  and b-quarks, and high resolution
calorimetry for jets and missing transverse energy $E^{miss}_{T}$.

The second supergoal of the LHC project is the supersymmetry discovery,
i.e. the detection of superparticles. Here the main
signature are the missing transverse energy events which are
the consequence of undetected lightest stable supersymmetric
particles LSP predicted in supersymmetric models with R-parity
conservation. Therefore it is necessary to set stringent requirements
for the hermeticity and $E^{miss}_{T}$ capability of the detector.
Also the search for new physics different from supersymmetry
(new gauge bosons $W^{'}$ and $Z^{'}$, new Higgs bosons with big
Yukawa couplings etc.) at LHC requires high resolution lepton
measurements and charge identification  even in
the $p_{T}$ range of a few TeV. Other possible signature of new
physics(compositeness) can be provided by very high $p_{T}$ jet
measurements. An important task of LHC is the study of b- and
t-physics. Even at low luminosities the LHC will be a high rate
beauty- and top-quark factory. The main emphasis in B-physics
is the precise measurement of CP-violation in the $B^{0}_{d}$
system and the determination of the Kobayashi-Maskawa angles. Besides,
investigations of $B\bar{B}$ mixing in the $B^{0}_{S}$ system, rare B
decays are also very important. Precise secondary vertex determination,
full reconstruction of final states with relatively low-$p_{T}$
particles, an example being $B^{0}_{d} \rightarrow J/\Psi K^{0}_{S}$
followed by $J/\Psi \rightarrow l^{+}l^{-}$ and $K^{0}_{S}
\rightarrow \pi^{+}\pi^{-}$, and low-$p_{T}$ lepton first-level
triggering capability are all necessary. In addition to running
as a proton-proton collider, LHC will be used to collide heavy
ions at a centre of mass energy 5.5 TeV per nucleon pair.
The formation of quark-gluon plasma in the heavy ion collisions
is predicted to be signalled by a strong suppression of
$\Upsilon^{'}$ and $\Upsilon^{''}$ production relative to
$\Upsilon$ production when compared with pp collisions.
The CMS and ATLAS detectors will be
used to detect low momentum muons produced in heavy ion collisions
and reconstruct $\Upsilon$, $\Upsilon^{'}$ and $\Upsilon^{''}$ meson
production. Therefore the basic design considerations for both ATLAS
and CMS are the following:

1. very good electromagnetic calorimetry for electron and photon
identification and measurements,

2. good hermetic jet and missing $E_{T}$-calorimetry,

3. efficient tracking at high luminosity for lepton momentum
measurements, for b-quark tagging, and for enhanced electron and photon
identification, as well as tau and heavy-flavour vertexing and
reconstruction capability of some B decay final states at lower
luminosity,

4. stand-alone, precision, muon-momentum measurement up to highest
luminosity, and very low-$p_{T}$ trigger capability at lower luminosity,

5. large acceptance in $\eta$ coverage.

\subsubsection{Brief description of CMS subdetectors \cite{67}}

The CMS detector consists of inner detector(tracker), 
electromagnetic calorimeter, hadron calorimeter, 
muon spectrometer and trigger. A schematic view of the CMS 
detector is shown in Fig.17.

\begin{figure}[hbt]

\vspace*{0.5cm}
\hspace*{0.0cm}
\epsfxsize=10cm \epsfbox{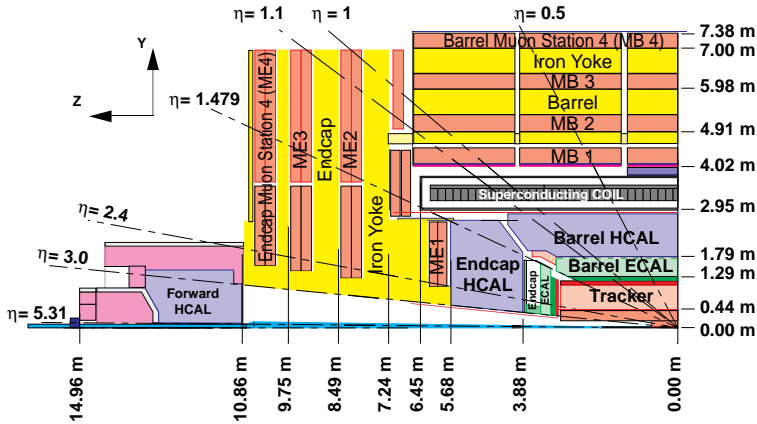}
\vspace*{0.0cm}
\caption[]{\label{fg:CMS} \it Longitudinal view of the CMS detector.}

\end{figure}

{\bf Tracker}

The design goal of the central tracking system is to reconstruct
isolated high $p_{T}$ tracks with an efficiency better than $95$
percent, and high $p_{T}$ tracks within jets with an efficiency of
better than 90 percent over the rapidity $|\eta| \le 2.6$. The
momentum resolution required for isolated charged leptons in the
central rapidity region is $\frac{\delta p_{T}}{p_{T}} =  0.1p_{T}$
$(p_{T}$ in $TeV$). This will allow the measurement of the lepton
charge up to $p_{T} = 2~TeV$. It is also very important for tracking
system to perform efficient $b$- and $\tau$-tagging.  The tracker
system consists of silicon pixels, silicon and gas microstrip
detectors(MSGS) which provide precision momentum measurements and
ensure efficient pattern of recognition even at the highest
luminosity. A silicon pixel detectors consist of two barrel layers
and three endcap layers and it is placed close to the beam pipe with
the tasks of:

a. assisting in pattern recognition by providing two or three
true space points per track over the full rapidity range in the
main tracker,

b. improving the impact parameter resolution for b-tagging,

c. allowing 3-dimensional vertex reconstruction by providing a much
improved $Z$-resolution in the barrel part.

The silicon microstrip detector is required to have a powerful vertex finding
capability in the transverse plane over a large momentum range for $b$-tagging
and heavy quark physics and must be able to distinguish different interaction
vertices at high luminosity. The CMS silicon microstrip detector is
subdivided into barrel and forward parts, meeting at $|\eta| = 1.8 (\eta
\equiv -\ln(\tan(\frac{\theta}{2}))$, provided at least 3 measuring
points on each track for $|\eta| \le 2.6$. The microstrip gas
chambers provide a minimum of 7 hits for high $p_{T}$ tracks. The
track finding efficiency in the tracker is 98 percent for $p_{T} \ge
5$ GeV. The charged particle momentum resolution depends on the
$\eta$ and $p_{T}$ of charged particle and for $p_{T} = 100~GeV$ and
$|\eta| \le 1.75$ it is around 2 percent. Impact parameter resolution
also depends on $p_{T}$ and $\eta$ and for $10 ~GeV \le p_{T} \le 100
 ~GeV$ and $|\eta| \le 1.3$ in transverse plane it is around 100
$\mu m$. The b-tagging efficiency from $\bar{t}t$ decays is supposed to be
better than 30 percent. A significant impact parameter 
resolution can be used to
tag $\tau$-leptons. It could be useful in searches such as SUSY Higgs
boson decays $A,H,h \rightarrow
\tau\tau \rightarrow e + \mu + X$(or $l +
hadrons)$.  These leptons(hadrons) originate from secondary $(\tau)$
vertices while in the backgrounds from $\bar{t}t \rightarrow Wb +
W\bar{b} \rightarrow e + \mu + X$ and
$WW \rightarrow e + \mu + X$ they originate
from the primary vertex. It is possible to have the
efficiency for the signal $\approx 50$ percent while for the
background channels it is $\approx 3$ percent.

{\bf ECAL}

The barrel part of the electromagnetic calorimeter covers the rapidity
intervals $|\eta| \le 1.56$. The endcaps cover the intervals
$1.65 \le |\eta| \le 2.61$. The gaps between the barrel and the
endcaps are used to route the services of the tracker and preshower
detectors.  The barrel granularity is 432 fold in $\phi$ and
$108 \times 2$-fold in $\eta$.  A very good intrinsic energy
resolution given by
\begin{equation}
\frac{\sigma}{E} =
\frac{0.02}{\sqrt{E}} \oplus 0.005 \oplus \frac{0.2}{E}
\end{equation}
is assumed to be for electrons and photons with a $PbWO_{4}$ crystal
ECAL. The physics process that imposes the strictest performance
requirements on the electromagnetic calorimeter is the intermediate
mass Higgs boson decaying into two photons. The main goal here is to obtain
very good di-photon mass resolution. The mass resolution has terms
that depend on the resolution in energy $(E_{1}, E_{2})$ and the two
photon angular separation $(\theta)$ and it is given by
\begin{equation}
\frac{\sigma_{M}}{M} = \frac{1}{2}[\frac{\sigma_{E_{1}}}{E_1} \oplus
\frac{\sigma_{E_2}}{E_2} \oplus
\frac{\sigma_{\theta}}{(\tan(\frac{\theta}{2})}] \,,
\end{equation}
where $\oplus$ denotes a quadratic sum, $E$ is in GeV
and $\theta$ is in radians. For the Higgs two-photon decay at LHC the angular
term in the mass resolution can become important, so it is necessary to
measure the direction of the photons using the information from the
calorimeter alone. In the barrel region $|\eta| \le 1.56$ angular
resolution is supposed to be $\sigma_{\theta} \le \frac{50
mrad}{\sqrt{E}}$.  Estimates give the following di-photon mass
resolution for $h \rightarrow \gamma \gamma$ channel $(m_{h} = 100
 ~GeV$):

$\delta m_{\gamma\gamma} = 475~MeV$ (Low luminosity
$L = 10^{33}~cm^{-2}s^{-1}$),

$\delta m_{\gamma\gamma} = 775 MeV$
(High luminosity $L =
10^{34}~cm^{-2}s^{-1}$).

{\bf HCAL.}

The hadron calorimeter surrounds the electromagnetic calorimeter and acts in
conjunction with it to measure the energies and directions of particle jets,
and to provide hermetic coverage for measurement the transverse energy. The
pseudorapidity range $( |\eta| \le 3)$ is covered by the barrel and
endcap hadron calorimeters which sit inside the $4T$ magnetic field of CMS
solenoid. In the central region around $\eta = 0$ a hadron shower
'tail catcher' is installed outside the solenoid coil to ensure
adequate sampling depth. The active elements of the barrel and endcap
hadron calorimeter consist of plastic scintillator tiles with
wave length-shifting fibre readout. The pseudorapidity range $( 3.0
\le \eta \le 5.0)$ is covered by a separate very forward
calorimeter. The hadron calorimeter must have good hermeticity, good
transverse granularity, moderate energy resolution and sufficient
depth for hadron shower containment. The physics programme requires
good hadron resolution and segmentation to detect narrow states
decaying into pairs of jets. The di-jet mass resolution includes
contributions from physics effects such as fragmentation as well as
detector effects such as angular and energy resolution. The energy
resolution is assumed to be:
\begin{equation}
\frac{\Delta E}{E} = \frac{0.6}{\sqrt{E}} \oplus 0.03
\end{equation}
for $|\eta| \le 1.5$ and segmentation $\Delta \eta \times \Delta \Phi
= 0.1 \times 0.1$.

The di-jet mass resolution is approximately the following:

1. $(10 -15)\%$ for $50 ~GeV \le p_{T} \le 60~GeV$ and $m_{ij} =m_Z$.

2. $(5 - 10)\%$  for $500 ~GeV \le p_{T} \le 600~GeV$ and $m_{ij} =m_Z$.

The expected energy resolution for jets in the very forward
calorimeter is parametrised by:
\begin{equation}
\frac{\sigma_{E_{jet}}}{E_{jet}} = \frac{1.28 \pm 0.1}{\sqrt{E_{jet}}}
\oplus (0.02 \pm 0.01)\,.
\end{equation}
The expected missing transverse energy resolution in the CMS detector with
very forward $2.5 \le \eta \le 4.7$ coverage is
\begin{equation}
\frac{\sigma_t}{\sum E_t} = \frac{0.55}{\sqrt{\sum E_t}} \,,
\end{equation}
($E_t$ in GeV). In the absence of the very forward calorimeter, the missing
transverse energy resolution would be nearly three times worse.

{\bf Muon system.}

At the LHC the effective detection of muons from Higgs bosons, $W$, $Z$ and
$t\bar{t}$ decays requires coverage over a large rapidity interval. Muons
from pp collisions are expected to provide clean signatures for a wide
range of new physics processes. Many of these processes are expected
to be rare and will require the highest luminosity. The goal of the muon
detector is to identify these muons and to provide a precision measurement
of their momenta from a few $GeV$ to a few $TeV$. The barrel detector covers
the region $|\eta| \le 1.3$. The endcap detector covers the region
$1.3 \le |\eta| \le 2.4$. The muon detector should fulfil three
basic tasks: muon identification, trigger and momentum measurement.
The muon detector is placed behind ECAL and the coil. It consists of
four muon stations interleaved with the iron return yoke plates. The
magnetic flux in the iron provides the  possibility of an independent
momentum measurement.  The barrel muon detector is based on a system
of 240 chambers of drift tubes arranged in four concentric stations.
In the endcap regions, the muon detector comprises four muon
stations. The muon detector has the following functionality and
performance:

1. Geometric coverage: pseudorapidity coverage up to $|\eta| =2.4$ with
the minimum possible acceptance loses due to gaps and dead areas.

2. Transverse momentum resolution for the muon detector alone for
$0 \le |\eta| \le 2$ : $\frac{\Delta p_{T}}{p_{T}} = 0.06 - 0.1$ for
$p_{T} = 10 ~GeV$, $0.07 - 0.2$ for $p_{T} = 100`GeV$ and $0.15 -
0.35$ for $p_{T} = 1 ~TeV$.

3. Transverse momentum resolution after matching with central detector
for $0 \le |\eta| \le 2$ : $\frac{\Delta p_{T}}{p_{T}} = 0.005 -
0.01$ for $p_{T} = 10 ~GeV$, $0.015 - 0.05$ for $p_{T} = 100 ~GeV$
and $0.05 - 0.2$ for $p_{T} = 1~TeV$.

4. Charge assignment: correct at 99 percent confidence level up to
  $p_{T} = 7$ TeV for the full $\eta$ coverage.

5. Muon trigger: precise muon chambers and fast dedicated detectors provide
 a trigger with $p_{T}$ thresholds from a few $GeV$ up to $100~GeV$.

{\bf Trigger.}

For the nominal LHC design luminosity $10^{34}~cm^{-2}s^{-1}$, an average 
of 20 inelastic events occur every $25~ns$, the beam crossing time interval. 
The input rate of $10^9$ interactions per second must be reduced by a 
factor of at least $10^{7}$ to $100~Hz$, which is the maximum rate that 
should be achieved for off-line analysis. CMS reduces this rate in 
two steps. The Level-1 trigger system operates on a subset of the data 
collected from each LHC crossing. The processing is dead timeless and the 
decision to collect the full set of data relating to a given crossing is 
taken after a fixed latency of $3~\mu s$. The maximum event rate which 
can be accepted by the Level-2 trigger, which again considers a subset of 
data, is $100~kHz$. The Level-1 trigger system comprises the front-end  
electronics which generates trigger primitives at the detector and the 
Level-1 processing logic in the electronic barracks, interconnected 
electrically and optically. The Level-2 trigger is provided by an online 
processor farm. After a Level-2 positive decision, the remainder of the 
full crossing data is requested for further processing by this farm 
for the final (Level 3) decision. 

The benchmarks for the trigger selection correspond to the final 
states which are not interesting in their own right, but  typical 
of final states expected in new physics processes. They correspond to 
inclusive triggers that must be highly efficient for new physics 
signatures. The benchmarks are:

1) electrons from inclusive $W$ bosons,

2) muons from inclusive $W$ bosons,

3) jets at high $p_t$, 

4) high $p_t$ photons,

5) missing $E_{T}$,

6) low $p_T$ multi leptons (for $b$ physics).

\subsection{ATLAS detector \cite{68}}

The design of the ATLAS detector is similar to CMS detector. It also consists
of inner detector(tracker), electromagnetic calorimeter, hadron calorimeter
 ,muon spectrometer and trigger.  Here we briefly describe the main 
parameters of the ATLAS subdetectors. A schematic view 
of the ATLAS detector is shown in Fig.18.

\begin{figure}[hbt]

\vspace*{0.5cm}
\hspace*{0.0cm}
\begin{turn}{-90}%
\epsfxsize=10cm \epsfbox{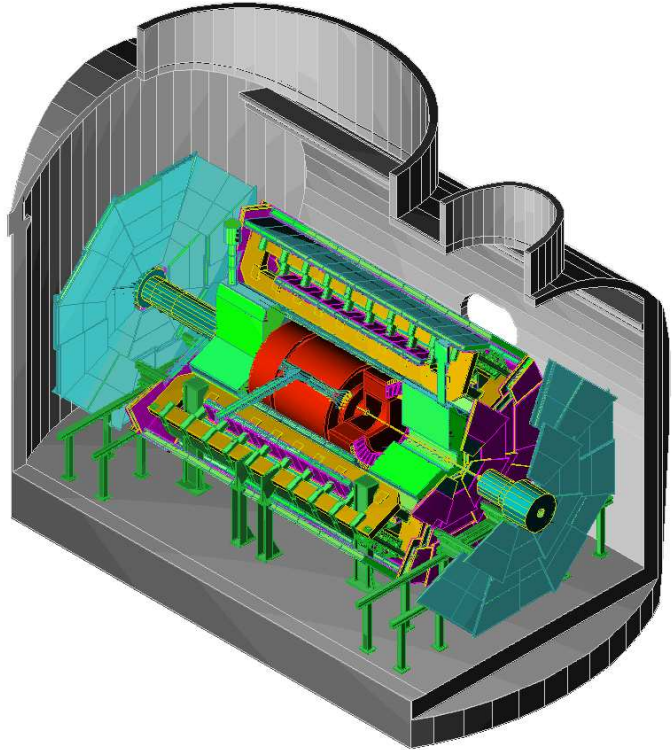}
\end{turn}
\vspace*{0.0cm}

\caption[]{\label{fg:ATLAS} \it Tree-dimensional view of the ATLAS detector.}
\end{figure}

{\bf Inner detector}

The main parameters of the ATLAS inner detector at high-luminosity 
running are:

1. Tracking coverage over the pseudorapidity range $|\eta| \le 2.5$.

2. Momentum resolution of $\frac{\Delta p_{T}}{p_T} \le 0.3$ at
$p_T = 500 ~GeV$  for $|\eta| \le 2$ and no worse than 50 percent for
$|\eta| = 2.5$.

3. Polar-angle resolution  of $\le 2$ mrad.

4. Tracking efficiency of $\geq 95 \% $  over the full coverage for
isolated tracks with $p_{T} \geq 5$ GeV, with fake-track rates less
than $1\% $  of signal rates.

5. Tracking efficiency of $\geq 90\, \%$ for all tracks with 
$p_T > 1~GeV$ in a cone $\Delta R < 0.25$ around high-$p_{T}$ 
isolated track candidates, with less than $10 \,\% $ of such tracks being 
fakes. Here, $\Delta R$ is defined as the separation of the 
particles in pseudorapidity-azimuth space. 

6. Electron-finding efficiency (integrated over all $p_{T}$, and including 
the trigger  efficiency) of $>90\,\% $ for a second electron    
with $p_T > 0.5~GeV$ near a high $p_T$-candidate, in order 
to suppress photon-conversion and Dalitz-decay backgrounds.

7. High $ p_{T} $-electron identification efficiency above $90\, \% $ 
both in the trigger and in the full reconstruction, including the effects
 
of bremsstrahlung in the tracker material. 

8. Combined efficiency of the calorimeter and inner detector in excess of 
 $85 \, \% $ for finding photons in the $p_T \sim 60 ~GeV $, with an 
electron rejection factor $>500$ and with an isolated $\pi^0$ rejection 
factor $>3$.  
 
9. Tagging of $b$ jets with an efficiency $\ge 30\, \% $  at the
highest luminosity, with a rejection $\ge 10$ against non $b$- hadronic
jets.

10. Measurement of the $z$ coordinate of primary vertices with at least 
four charged tracks to better than $1~mm$.

11. Provision of LVL2 track trigger to select isolated tracks with 
$p_T > 20~GeV$ with an efficiency $>90\,\%$ and a fake track rate of 
$<10\,\%$, in a cone of $\Delta R < 0.25$ around high-$E_T$ e.m. 
calorimeter clusters.  

For initial lower-luminosity running the additional important parameters are:
 
1. Tagging of $b$-jets with an efficiency above $30\,\% $, with a rejection 
$>50$ against non $b$-hadronic jets. 

2. The ability to reconstruct secondary
vertices from $b$ and $\tau$ decays and charged tracks from primary
vertices and from secondary decay vertices
of short-lived particles
with $\ge 95 \, \% $  efficiency for $p_T \ge 0.5~GeV$ over the full
coverage.

3. Reconstruction and identification of electrons with $p_T > 1~GeV$ 
with an efficiency $>70\,\% $.

{\bf ECAL}

The energy resolution is of $ \frac{\Delta E}{E} =
\frac{0.1}{\sqrt{E}} \oplus 0.007$ for $|\eta| \le 2.5$. Diphoton
mass resolution is estimated to be $1.4 ~GeV$ for Higgs boson mass $m_h
= 100~GeV$ for $L = 10^{34}~cm^{-2}s^{-1}$ (for CMS the diphoton mass
resolution is $775~MeV$).

{\bf HCAL}

Jet energy resolution is of $\frac{\Delta E}{E} = \frac{0.5}{\sqrt{E}}
\oplus 0.03$ for jets and a segmentation of $\Delta \eta \times
\Delta \Phi = 0.1 \times 0.1$ for $|\eta| \le 3$ and
$\frac{\Delta E}{E} = \frac{1}{\sqrt{E}} \oplus 0.1$ and a segmentation
of $\Delta \eta \times \Delta \Phi = 0.1 \times 0.1$ for very forward
calorimeter $3 \le |\eta| \le 5$.

{\bf Muon spectrometer}

The muon momentum resolution is of $\frac{\Delta p_{T}}{p_T} = 0.02
(p_T = 20 ~GeV$), $\frac{\Delta p_T}{p_T} = 0.02(p_T = 100~GeV$),
$\frac{\Delta p_{T}}{p_T} = 0.08(p_T = 1~TeV$) for $|\eta| \le 3$.

{\bf Trigger}

The ATLAS trigger is organised in three trigger levels (LVL1, LVL2, LVL3). 
At LVL1, special- purpose processors act on reduced-granularity data from 
a subset of the detectors. The LVL2 trigger uses full-granularity, 
full-precision data from most of the detectors, but examines only regions 
of the detector identified by LVL1 as containing interesting information. 
At LVL3, the full event data are used to make the final selection of 
events to be recorded for offline analysis. The LVL1 trigger accepts 
data at the full LHC bunch-crossing rate of 40 MHz (every 25 ns) and 
reduces them to 100 Khz. The LVL2 trigger reduces the rate from 
up to 100 KHz after LVL1 to about 1 KHz. After an event is accepted 
by the LVL2 trigger, the full data are sent to the LVL3 processors which 
must achieve a data-storage rate of $10 -100 ~MB/s$ by reducing 
the event rate and/or the event size.

\section{ Search for standard Higgs boson at the 
LHC}

In this section we give mainly the results of the simulations on 
the search for Higgs boson at CMS detector \cite{67},  
\cite{70} -\cite{86}. We don't give the review of the corresponding 
ATLAS simulations \cite{68}, \cite{87} -\cite{96} on the Higgs boson 
search because the results in terms of the significances 
coincide up to $30 \%$.  However sometimes we  compare the CMS 
and ATLAS Higgs boson discovery potentials.

\subsection{The search for $h \rightarrow \gamma \gamma$.}

One of the most important reactions for the search for Higgs boson at
LHC is
\begin{equation}
pp \rightarrow (h \rightarrow \gamma\gamma) +...\,,
\end{equation}
which is the most promising one for the search for Higgs boson in the
most interesting region $100~GeV \le m_{h} \le 140~GeV$.

The key features that enable CMS detector to obtain clear two-photon
mass peaks, significantly above background throughout the
intermediate mass range, are:

i. An electromagnetic calorimeter with an excellent energy resolution
(this requires calibration to high precision, which in turn requires
a good inner tracking system).

ii. A large acceptance (the precision electromagnetic calorimetry
extends to $|\eta| = 2.5$), adequate neutral pion rejection and (at
high luminosity) a good measurement of photon direction. This
requires fine lateral segmentation and a preshower detector.

iii. Use of powerful inner tracking system for isolation cuts.

The cross section ( including K-factor $K = 1.5$) 
times branching has been estimated to be
$\sigma Br(h \rightarrow \gamma \gamma ) = 76~fb(68~fb)$
for $m_h =110(130)~GeV$, the uncertainty in
the cross section calculation is
(10 - 30) percent. The imposition of cuts ($|\eta| \le 2.5$,
$p_{T}^{\gamma_1} \geq 40 ~GeV$, $p_{T}^{\gamma_2} \geq 25~GeV$)
allow to decrease the background in a reasonable magnitude. The jet
background is reduced by imposing an isolation cut, which also
reduces the bremsstrahlung background. Photon is defined to be
isolated if there is no charged track or electromagnetic shower
with a momentum greater than $2.5 ~GeV$ within a region $\Delta R
\leq 0.3$ around it. The photons from the decay of $\pi^{0}$ of the
relevant transverse momenta are separated in the calorimeter by a
lateral distance of the order of 1 cm. An efficiency of $64\,\% $ was
assumed for reconstruction of each photon (i.e. $41 \,\% $ per
event). The crystal calorimeter was assumed to have an energy
resolution $\Delta E/E = 0.02/\sqrt{E} \oplus 0.005 \oplus 0.2/E$ in
the barrel and $\Delta E/E = 0.05/\sqrt{E} \oplus 0.005 \oplus 0.2/E$
in the endcap, where there is a preshower detector. At high
luminosity, a barrel pre-shower detector covers $|\eta| < 1.1$,
resulting in  a resolution $ \Delta E/E = 0.05/\sqrt{E} \oplus 0.005
\oplus 0.2/E$ and an ability to measure the photon direction with
resolution $\Delta \alpha = 40$ $mrad/\sqrt{E}$ in this region.

The background to the $h \rightarrow \gamma \gamma$ may  be divided
into 3 categories:

1. prompt diphoton production from quark annihilation and gluon
 fusion diagrams - irreducible background,

2. prompt diphoton production from bremsstrahlung from the outgoing
 quark line in the QCD Compton diagram,

 3. background from jets, where an electromagnetic energy deposit
 originates from the decay of neutral hadrons in a jet from 1 jet +
 1 prompt photon.

 The signal  significance $\sigma =
 \frac{N_S}{\sqrt{N_B}}$ is estimated to be $6.6\sigma(9\sigma)$ for
 $m_h = 110(130~GeV$ and for low luminosity $L_{low,t} = 3\cdot
 10^{4}~pb^{-1}$ and $10\sigma(13\sigma)$ for $m_h = 110(130) ~GeV$ and
 for high luminosity $L_{high,t} = 10^{5}~pb^{-1}$. The general
 conclusion is that at $5\sigma$ level it would be possible to discover
 Higgs boson \footnote{It should be noted that more correct definition 
of the significance in future experiments when we know only the 
average number of signal $N_S$ and background  $N_B$ events is 
$S = \sqrt{N_S +N_B} - \sqrt{N_B}$ \cite{97}. More appropriate 
characteristic for future experiments is the probability of the discovery, 
i.e. the probability that future experiment will measure the number 
of events $N_{ev}$ such that the probability that standard physics 
reproduces $N_{ev}$ is less than $5.7 \cdot10^{-7}$ ($5\sigma$). For instance, 
for the standard Higgs boson search with $m_h = 110~GeV$ and 
for $L = 3\cdot 10^{4}~pb^{-1}(2\cdot10^{4}~pb^{-1})$ the standard 
significance is $6.6(5.4)$. At the language of the probabilities it 
means \cite{97} that the CMS will discover at $\geq 5\sigma$ the 
Higgs boson with the probability $96(73)$ percent.}       
for $95 ~GeV \le m_{h} \le 145 ~GeV$ at low 
luminosity and at high
luminosity the corresponding Higgs boson mass discovery interval is
$85 ~GeV \le m_{h} \le 150~GeV$ (see Fig.19).

\begin{figure}[hbt]

\vspace*{0.5cm}
\hspace*{0.0cm}
\epsfxsize=10cm \epsfbox{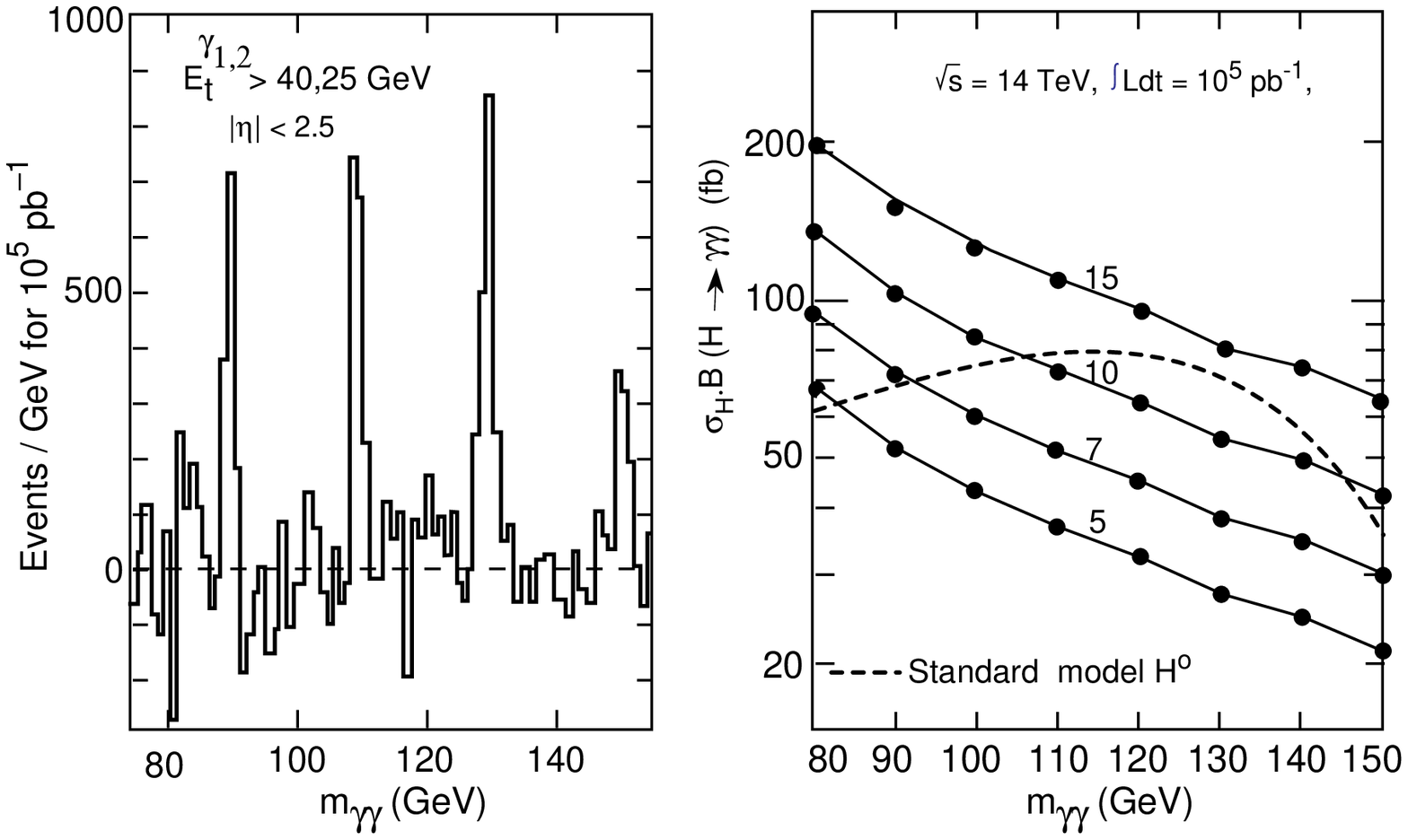}
\vspace*{0.0cm}
\caption[]{\label{fg:2gamma} \it (a) Background-subtracted 2 
$\gamma$ mass plot for $10^5pb^{-1}$ with signals at 
$m_h=~90,~110,~130$ and $150~GeV$ in $PbWO_4$ calorimeter (CMS).

(b) Signal significance contours for $10^5pb^{-1}$ taken at high
luminosity (CMS) ($H \equiv h$).}
\end{figure}

Comparison of the ATLAS and CMS discovery potential for the 
$h \rightarrow \gamma \gamma$ channel  has been made in ref. \cite{89}. 
The ratio between the CMS and ATLAS significances is determined by 
the formula
\begin{equation}
\frac{S_{CMS}}{S_{ATLAS}} \approx \sqrt{(\frac {\Delta_m(ATLAS)}
{\Delta_m(CMS)})} \times \frac{\epsilon_{\gamma}(CMS)}{\epsilon_{\gamma}
(ATLAS)}\,,
\end{equation}
where $\Delta_m$,  the diphoton mass resolution and $\epsilon_{\gamma}$, 
 the total photon efficiency(trigger, identification, reconstruction) 
are detector dependent. In ATLAS and CMS the photon identification 
efficiencies are  $80 \%$ \cite{68} and $71 \%$ \cite{67} correspondingly.    
However the diphoton mass resolution is better in CMS. According to 
the Technical Proposals for $m_h = 110~GeV$ the diphoton mass resolutions are:

$$\Delta_m(CMS) = 0.54~GeV, \: \Delta_m(ATLAS) =1.25~GeV\:(low~luminosity)$$

$$\Delta_m(CMS) = 0.87~GeV, \: \Delta_m(ATLAS) = 1.43~GeV\:(high~luminosity)$$

The main conclusion of the ref.\cite{89} is that the discovery potential 
of the CMS (in terms of $\sigma$) is $10\,\%$ and $30\,\%$ better than 
ATLAS at high and low 
luminosities stagies correspondingly (for $m_h =110~GeV$).

\subsection{ Search for $h \rightarrow \gamma\gamma$ in association
with high $E_{T}$ jets.}

The idea to look for Higgs boson signal associated with a high $p_t$ 
jet in the final state was considered in ref.\cite{82}, where the 
matrix elements of signal subprocesses $gg \rightarrow g + h$, 
$gq \rightarrow q + h$ and $q\bar{q} g + h$ have been calculated 
analytically in the leading order $\alpha^3_s$. One 
kind of reducible  background 
comes from the  reactions $qg \rightarrow \gamma +g +q$, $gg \rightarrow 
\gamma +q + \bar{q}$, $qq^{'} \rightarrow \gamma + q(g) + 
q^{'}(g)$ in the cases when the final gluon or quark produces an energetic 
photon without further jet generation. Other kind of reducible 
background comes from the subprocesses 
$qg \rightarrow \gamma + q, q\bar{q} \gamma + g$ when the second 
photon is produced during the quark or gluon fragmentation but this 
jet is still detected. Third kind of reducible background could come 
from the pure QCD subprocesses $2 \rightarrow 2$ type, when both 
particles in the final state  are gluons and quarks. There is nonzero 
probability to get two separated and energetic photons from the 
fragmentating of quarks and gluons. There are possible contributions 
from the following subprocesses: $gg \rightarrow g(q) + q(\bar{q})$, $gq 
\rightarrow g +q $, $qq^{'}  \rightarrow q(g) + q^{'}(g)$.

The typical set of cuts used to separate signal from background is \cite{82}:

(C1) two photons are required with $p^{\gamma}_t > 40 ~GeV$, and 
$|\eta|_{\gamma} < 2.5$ for each photon;

(C2) photons are isolated from each other by 
$\Delta R(\gamma_1, \gamma_2) > 0.3$;

(C3) jet has high transverse energy $E^{jet}_t > 40 ~GeV$ and is 
centrally produced, $|\eta_{jet}| < 2.4$;

(C4) jet is isolated from the photons by $\Delta R(jet, \gamma_1) >0.3$ 
and $\Delta R(jet, \gamma2) > 0.3$.  

For the Higgs boson mass 
$ 100 ~GeV \leq M_h \leq 150 ~GeV$ and for an integrated luminosity 
$10 ~fb^{-1}$ this channel has dozens  of signal events with a number of 
background events only by factor 2-3 higher \cite{82}. The significance 
$N_S/\sqrt{N_B} \sim 4.0; 5.3$ and 4.1 for $M_h = 100, 120 $ and $140~GeV$ 
respectively indicating good prospects for discovery of the light 
Higgs boson at low LHC luminosity. These result also imply that 
at high luminosity phase with year luminosity $10^5 ~pb^{-1}$ LHC will 
give hundred of events with high $p_t$ associated with hard jet with the 
signal significance $\sim 15$. 

Note that recent study \cite{92} of the signature $\gamma \gamma + ~jets$ 
for ATLAS detector confirms the main results 
of ref.\cite{82}.

The possibility for the search for $h
\rightarrow \gamma\gamma$ with $ \geq 2$ large $E_{T}$ jet 
also allows to improve
 $signal/background$ ratio. There are several sources of such Higgs +
 jet events. One is the next to leading order corrections to $gg
 \rightarrow h$ with hard gluons.  Others are the associated
 production of $\bar{t}th, Wh, Zh$ and the WW and Zh fusion
 mechanisms.

The cuts that provide optimal sensitivity are \cite{73}:

i. Two isolated photons are required, with $p_t^{\gamma_{1}}
\geq 40 ~GeV$ and $p_{t}^{\gamma_{2}} \geq 60~GeV$, $|\eta| \leq 2.5$ and
$p_t^{\gamma \gamma} \geq 50~GeV$.

ii. Number of jets $\geq 2$, $E_t^{jet} \geq 40~GeV$ for the
central jets ( $|\eta| \leq 2.4$ ) and $E^{jet} \geq 800~GeV$
for the forward ones ($2.4 \leq |\eta| \leq 4.6$).

iii. Photons are isolated with no charged or neutral particles with
$p_t \geq 2 ~GeV$ within a cone $\Delta R \leq 0.3$ around each
photon's direction.

iiii. $\gamma$-jet isolation $\Delta R(\gamma ,jet) > 1.5$ (to
suppress the bremsstrahlung contribution).

The calculations give encouraging results, namely for
 $L_{high,t} = 1.6\cdot10^{5}~pb^{-1}$ it would be possible to
 discover the Higgs boson for $70 ~GeV \le m_h \le 150 ~GeV$ with $\ge
 7\sigma$ signal significance. Note that  the background is not only
much smaller in magnitude than in the inclusive $h \rightarrow
\gamma \gamma $ search, but it is also peaked at higher masses, away
from the most difficult region $m(\gamma \gamma) \leq 90 ~GeV$.

\subsection{$h \rightarrow W^+W^- \rightarrow l^+\nu l^- \nu$ signature.}

Recently it has been shown \cite{83} that the previously ignored signature 
$pp \rightarrow h \rightarrow W^+W^- \rightarrow l^+\nu {l^{'}}^-
\bar{\nu^{'}}$ 
provides the Higgs boson discovery for the Higgs boson mass 
region between $155 ~GeV$ and $180 ~GeV$ at the LHC. The proposed signature 
does not require extraordinary detector performance and only requires a 
relatively low integrated luminosity of about $5 ~fb^{-1}$. 

The main background production reactions are 
\begin{equation}
pp \rightarrow (W^+W^-, W^{\pm}Z^0, t\bar{t}, W^{\pm}t(b) + ...)
\end{equation}

The most important selection criteria for the enhancement of the signal 
over the background are the following \cite{83}.

1. Events which contain two isolated high $p_t$ charged leptons, 
electrons or muons, which are inconsistent with Z decays are selected. 
Both leptons  should have a pseudorapidity $|\eta|$ of less than 2.4 and 
their $p_t$ should be larger than $25~GeV$ and $10~GeV$ respectively. 
The dilepton mass should be larger than $10~GeV$ and more than 
$5 ~GeV$ 
different from Z boson mass if the event consists of $e^+e^-$ or 
$\mu^+\mu^-$ pairs. 

2. Background from $t\bar{t} \rightarrow bW^+\bar{b}W^-$ and 
$gb \rightarrow Wtb \rightarrow WbW(b)$is reduced by vetoing 
events which contain jets with $p_t$ of more than $20 ~GeV$ and 
$|\eta| <3 $ .

3. Signal events from gluon-gluon scattering are more 
central than the $W^+W^-$ background from $q\bar{q}$ scattering. 
This criterium is essentially independent of the mass. Therefore it 
is required that the polar angle of the reconstructed dilepton momentum 
vector, with respect to the beam direction, is larger than 30 degrees and 
that the absolute value of the pseudorapidity difference of the leptons 
is smaller than 1.25. As  a result both leptons are found essentially 
within the barrel region of the experiments with $|\eta|  < 1.5$.

4. The $W^+W^-$ spin correlations and the $V-A$ structure of the W decays 
result in a distinctive signature for $W^+W^-$ pairs produced in Higgs 
boson decays. For Higgs boson mass close to $2 \times M_W$ 
the $W^{\pm}$ boost is small and the opening angle between the two 
charged leptons in the plane transverse to the   beam direction 
is small. 

The results of the analysis \cite{83} demonstrate that this 
signature provides not only the Higgs boson discovery channel 
for a mass range between $(155 - 180)~GeV$ with $S/B \geq 0.35$ 
but also helps to establish a LHC Higgs boson signal for masses between 
$(120 -500)~GeV$. Recent simulations study \cite{84} based on PYTHIA to 
generate events and CMSCIM calorimeter simulation for the jet veto 
confirm qualitatively the results of ref.\cite{83}. Numerically they give 
for $m_h = 130~GeV$ $30 \,\%$ lower efficiencies for the signal and 
DY background and about a factor 2 higher overall efficiency for $t\bar{t}$ 
background \cite{84}.

\subsection{$h \rightarrow ZZ^{*}(ZZ) \rightarrow 4$ leptons.}

{\bf $m_h < 2m_Z$ region}

The channel $h \rightarrow ZZ^{*} \rightarrow 4\,l$ is the most  promising 
one to observe Higgs boson in the mass range $130~GeV - 180~GeV$.  
Below  $2M_Z$    the event rate is small and the background
reduction more difficult, as one of the $Z$s is off mass shell. In
this mass region the width of the Higgs boson is small $\Gamma_{h}
< 1~GeV$, and the observed width is  entirely determined by the 
instrumental mass resolution.    
 The significance of the signal is proportional to the
four-lepton mass resolution ($S = N_S/\sqrt{N_B}$ and 
$N_B \sim \sigma_{4l}$, so the lepton 
energy/momentum resolution is of decisive importance 
\footnote{ Typical Higgs boson mass resolutions in this mass range are: 
$\sigma_{4\mu} \approx 1~GeV,\, \sigma_{4e} \approx 1.5 GeV$ (CMS) \cite{67} 
and $\sigma_{4\mu} \approx 1.6 ~GeV$, $\sigma_{4e} \approx 1.6~GeV$ (ATLAS) 
\cite{68}. The comparison of the CMS and ATLAS discovery potentials 
with $h \rightarrow ZZ^{*} \rightarrow 4~leptons$ 
based on the analyses presented  in the two Technical Proposals 
has been performed in ref.\cite{88}. The main conclusion of the ref.\cite{88} 
is that in terms of significances ATLAS and CMS discovery potentials coincide 
up to $30, \% $.}.     
 
In the $m_h <2M_Z$ mass region , the main backgrounds are from
$t\bar{t}$, $Zb\bar{b}$ and $ZZ^{*}$. The $ZZ^{*}$ background is
irreducible and peaks sharply near the $ZZ$ threshold. The
$Zb\bar{b}$ background cannot be reduced by a $Z$ mass cut, but it
can be suppressed by lepton isolation. The $t\bar{t}$ background
can be reduced by a $Z$ mass cut and by isolation cuts. The standard
event cuts in CMS were chosen the following \cite{67}:
one electron with $p_t > 20~GeV$; one with $p_t > 15~GeV$, and the
remaining two electrons with $p_t >10~GeV$, all within $|\eta| <
2.5$. For muons, the corresponding $p_t$ cuts are 20, 10 and 5 $GeV$ in
the rapidity range $|\eta| <2.4$. For $m_h = 130~GeV$ the overall
(kinematic and geometrical) acceptance for the four-electron channel
is $22\,\% $ and for the four-muon channel $42 \, \% $. For $m_h =
170~GeV$ these acceptances increase to $38\,\% $ and $48\,\% $
respectively. To select $h \rightarrow ZZ^{*}$ events and
suppress the large $t\bar{t}$ background, one of the $e^{+}e^{-}$
or $\mu^{+}\mu^{-}$ pairs was assumed to be within $\pm 2\sigma_{Z}$
of the $Z$ mass. There is a fraction of events where both $Z$s are
off-shell. This effect results in a $24\,\% $ loss for $m_h = 130
~GeV$, decreasing to $12 \,\% $ for $m_h = 170 ~GeV$. The $M_Z$ cut
reduces $t\bar{t}$ background by a factor 11 in the $Z \rightarrow
\mu^{+}\mu^{-}$ channel and by a factor of 5 in the $Z \rightarrow
e^{+}e^{-}$ channel. For  two softer leptons, $M(ll) > 12$ GeV
is also required.
One can say that for the region $130 ~GeV \le m_{h} \le 180 ~GeV$ 
and  for $L_{high,t} = 10^{5}~pb^{-1}$ CMS will  discover the Higgs boson
with $\ge 5\sigma$ signal significance (see Fig.20) except 
narrow mass region around $170~GeV$ where $\sigma \times Br$ 
has a minimum due to the opening 
of the $h \rightarrow WW$ channel and drop of the 
$h \rightarrow ZZ^{*}$ branching ratio just below the $ZZ$ threshold.  
Note that the imposition of the additional cut on the mass of the 
second (lighter) lepton pair to $m_{34} < 76~GeV$ leads to a considerable 
signal improvement in this critical region.
At low luminosity $L = 2\cdot 10^{4}~pb^{-1}$ Higgs 
boson can be discovered at CMS in the mass range $m_h = (130 - 150)~GeV$.

\begin{figure}[hbt]

\vspace*{0.5cm}
\hspace*{0.0cm}
\epsfxsize=10cm \epsfbox{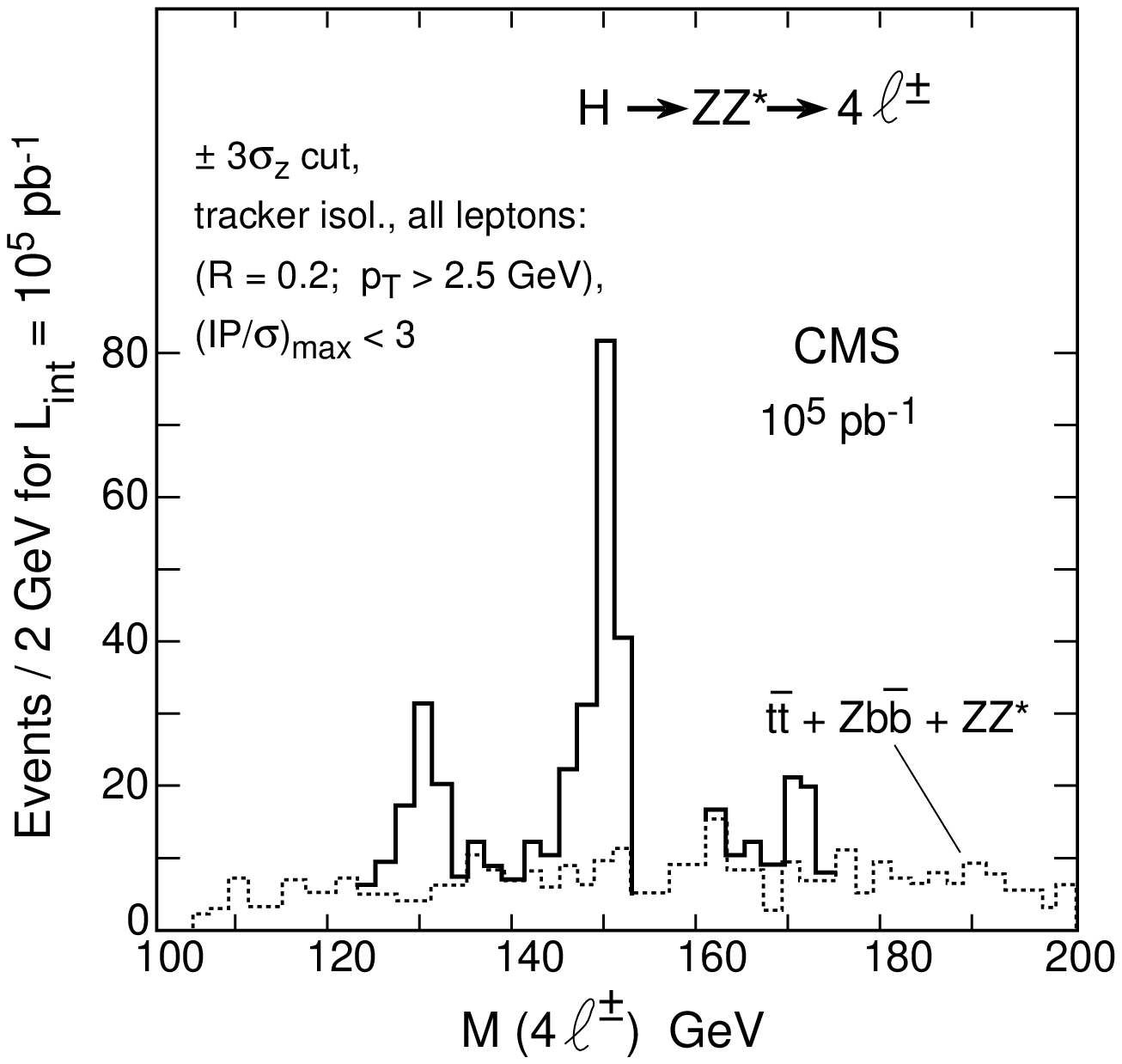}
\vspace*{0.0cm}

\caption[] {\label{fg:ZZstar} \it The four-lepton mass distributions for 
$h \rightarrow ZZ^* \rightarrow 4 l^{\pm}$ in CMS, superimposed
on the total background, for $m_h~=~130,~150$ and $170~GeV$ with
$10^5pb^{-1}$ ($H \equiv h$).}
\end{figure}

{\bf $h \rightarrow ZZ \rightarrow  4\,l$ }

For $180~GeV \leq m_h \leq 800~GeV$, this signature is considered to be 
the most reliable one for the Higgs boson discovery at LHC, since the 
expected signal rates are large and the background is stall.
The main background to the $h \rightarrow ZZ \rightarrow  4l^{\pm}$ 
process is the irreducible $ZZ$ production from $q\bar{q} \rightarrow ZZ$ 
and $gg \rightarrow ZZ$. The $t\bar{t}$ and $Zb\bar{b}$ backgrounds are  
small and reducible by a $Z$-mass cut. The typical cuts are the 
following \cite{76,77}: 

1. One electron with $p_T > 20~GeV$, one with $p_T  > 15~GeV$, and 
the remaining two electrons with $p_T> 10~GeV$, all within $|\eta| < 2.5$. 

2. For muons the corresponding $p_T$ cuts are 20,10 and 5 $GeV$, 
and the rapidity coverage is $|\eta| < 2.4$.

3. To avoid any residual $t\bar{t}$ background a cut on the $Z$-mass by 
$m_{l^+l^-} =m_Z \pm  4\sigma_Z$, with $\sigma_Z = 3~GeV$ is used.   

The use of the above determined cuts allows to detect the Higgs boson 
at $\geq 5\sigma$ level up to $\approx 400~GeV$ 
at $10^4~pb^{-1}$ and up to $m_h \approx 650~GeV$ at $10^{5}~pb^{-1}$ 
\cite{77} (see Fig.21). As it has been demonstrated in 
ref.\cite{77} the imposition of the additional cut 
$p^{Z_1}_T + p^{Z_2}_T > m_{ZZ}/1.4$ allows 
to extend the CMS discovery potential up to $650~GeV(3\cdot10^4~pb^{-1})$, 
 $750~GeV(10^{5}~pb^{-1})$, $850~GeV(3\cdot10^{5}~pb^{-1})$. 
  
Similar results have been obtained for ATLAS \cite{91}.

\begin{figure}[hbt]

\vspace*{0.5cm}
\hspace*{0.0cm}
\epsfxsize=10cm \epsfbox{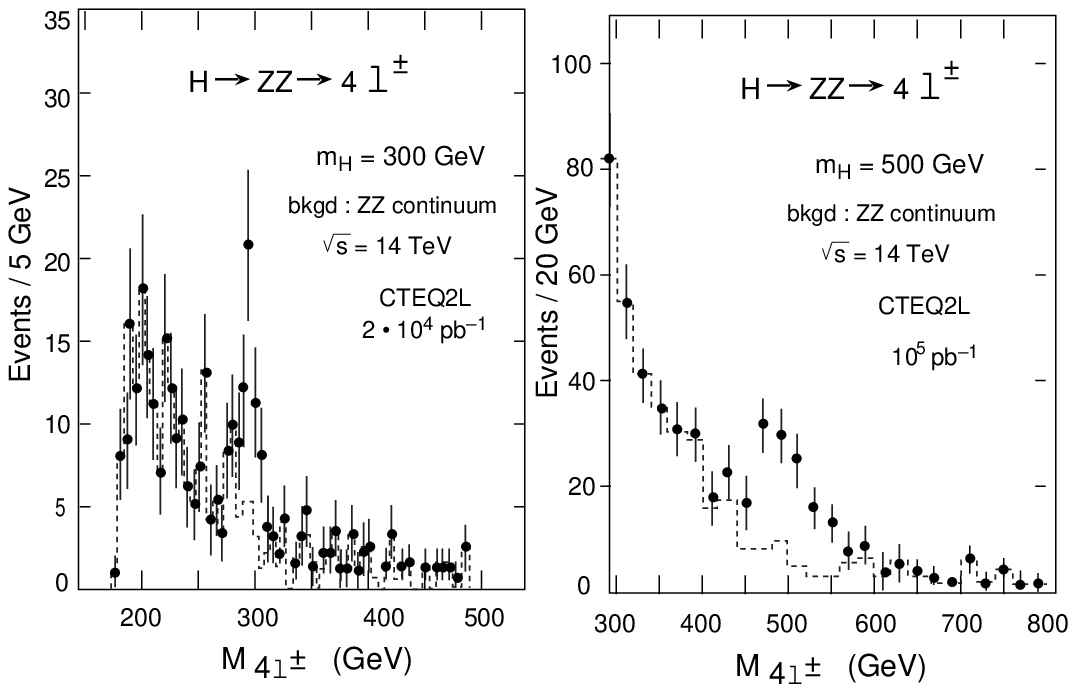}
\vspace*{0.0cm}

\caption[] {\label{fg:ZZ} \it The four-lepton mass distributions for 
$h \rightarrow ZZ \rightarrow 4 l^{\pm}$ in CMS, superimposed
on the $ZZ$ continuum background, for $m_h~=~300~GeV$ with
$2 \cdot 10^4pb^{-1}$ and for $m_h~=~500~GeV$ with
$10^5pb^{-1}$ ($H \equiv h$).}
\end{figure}


\subsection{The use of the signature $pp \rightarrow \gamma \gamma + 
lepton$}

The $Wh \rightarrow
l\gamma\gamma + X $ and $\bar{t}th \rightarrow l\gamma\gamma + X$
final states are other promising signature for the Higgs boson
search. The production cross section is smaller than the inclusive
$h \rightarrow \gamma \gamma $ by a factor $\approx 30$. However
the isolated hard lepton from the W and t decays allows to obtain a
strong background reduction and to indicate the primary vertex at
any luminosity.

Typical choice of cuts is the following \cite{67,72}:

1. $p_t^{\gamma_1} > 40~GeV, p_t^{\gamma_2} > 20~GeV$, transverse momentum 
cuts for photons.

2. $p^l_t > 20~GeV$, transverse momentum cuts for electron (or muon).

3. $|\eta_{\gamma_1,\gamma_2}| <2.4, |\eta_l| <2.4$, rapidity cuts for both 
photons and electron (or muon).

4. $\Delta R(\gamma_1, \gamma_2) > 0.3$ , $\Delta R(\gamma, l) >0.3$, 
isolation cuts for photon or photon-lepton pair. 

Here $\Delta R = \sqrt{\delta \phi^2 + 
\delta y^2}$ is the separation between two particles in the $\phi-y$ plane. 
The main background comes from the reactions $pp \rightarrow \gamma \gamma 
+ t\bar{t}$, $pp \rightarrow \gamma q(\bar{q} +  e^{\pm} \nu)$, 
$pp \rightarrow \gamma g + e^{\pm} \nu$ with a gluon jet misidentified 
as a photon, $pp \rightarrow \gamma + e^+e^-$ with an electron misidentified 
as a photon, $pp \rightarrow  
\gamma \gamma b\bar{b}$ with a b-quark misidentified as an 
electron.

The main conclusion is that  for an integrated luminosity 
$165 fb^{-1}$ in both channels $pp \rightarrow Wh$ and 
$pp \rightarrow t\bar{t}h$ in the two-photon invariant mass interval 
$M_h - 1 ~GeV \leq M_{\gamma \gamma} \leq M_h + 1~GeV$ there 
are $\sim 100$ signal for $M_h = 120~GeV$ 
and $\sim 20$ irreducible background events if the photon transverse 
momentum cuts are $20~GeV$. If the photon transverse momentum cuts are taken 
to be $40~GeV$ there are $\sim 50$ signal events and $1-2$ background events. 
Higgs peak can be observed practically free from the background. However in 
the low luminosity regime the reaction $pp \rightarrow \gamma \gamma + 
lepton$ is able to produce only $4-5$ clean signal events. So only in the 
high luminosity phase it allows to make an important cross-checking if 
the Higgs signal has shown up before in $pp \rightarrow h + ...
\rightarrow \gamma \gamma +...$ classical signature.

\subsection{The use of channels $h \rightarrow WW \rightarrow ll\nu\nu$,
$h \rightarrow WW \rightarrow l \nu jj$ and $h \rightarrow ZZ
\rightarrow ll jj$.}

The channel $h \rightarrow l l \nu \nu$ has a six times larger
branching than $h \rightarrow 4l^{\pm}$. The main background comes
from $ZZ$, $ZW$, $t\bar{t}$ and $Z + jets$.  The chosen cuts
are the following \cite{67}:

1. $E_t^{miss} \geq 100~GeV$.

2. Two isolated leptons are required, with $p_t \geq 20~GeV$, $|\eta|
\leq 1.8$ and $p_t^{ll} \geq 60$ GeV.

3. $|M_Z - M_{ll}| \leq 6~GeV$.

4. No other isolated leptons with $p_t \geq 6~GeV$.

5. No central jets with $E_t \geq 150~GeV$.

6. No jets back-to-back with leptons (cosine of the angle between
the momentum of the lepton pair and sum of the momenta of the jets
is  $\geq -0.8$).

 7. $E_t^{miss}$ vector back-to-back with the lepton pair (cosine
 of the angle in the transverse plane between the two-lepton momentum
 and the missing transverse momentum $\leq 0.8$).

The conclusion \cite{67,86} is that using this mode it 
would be possible to
discover Higgs boson in the interval $400 ~GeV
\le m_{h} \le (800 - 900) ~GeV$ (see Figs. 22-23).

The channels $h \rightarrow WW \rightarrow l\nu jj$ and
$h \rightarrow ZZ \rightarrow ll jj$ are important in the
$m_h \approx 1$ TeV mass range, where the large $W,Z
\rightarrow q \bar{q}$ branching ratios must be used.
Also high lepton pairs  with $m_{ll}  \approx M_Z$ for
$h \rightarrow ZZ$ or a high $p_t$ lepton pair
plus large $E^{miss}_t$, for $h \rightarrow WW$
must be used. In addition,
two hard jets from the hadronic decays of $Z/W$ with
$m_{jj}  \approx M_{Z/W}$ are required. The backgrounds are:
$Z + jets$, $ZW$, $WW$, $t\bar{t}$, $WW$, $WZ$. For
$m_h \approx 1 ~TeV$  the Higgs boson is very broad ($\Gamma_{h}
\approx 0.5~TeV$ and $WW/ZZ$ fusion mechanism represents about
50 percent of the total production cross section), therefore
forward-region signature is essential.
The appropriate cuts are the following:

i. $E_t^{miss} \geq 150 ~GeV$, $p^l_t \geq 150~GeV$, $p_t^W \geq 300
~GeV$ for $h \rightarrow WW$, or $p_t^l \geq 50 ~GeV$, $p_t^Z \geq 50
~GeV$, $p_t^Z \geq 150 ~GeV$, $|m_{Z} - m_{ll}| \leq 10 ~GeV$ for $h
\rightarrow ZZ$.

ii. $|m_{jj} - m _{W/Z}| \leq 15~GeV$ for the central jet pair.

iii. $E_t^{jet} \geq 10~GeV$, $E^{jet} \geq 400~GeV$, $|\eta|
\geq 2.4$ for the two forward tagging jets.

The main conclusion \cite{67, 78} is that the use 
of the reactions $h \rightarrow WW \rightarrow l\nu jj$ 
and $h \rightarrow ZZ \rightarrow lljj$ allows to discover the heavy 
Higgs boson with a mass up to $1~TeV$ for $L_{high,t} = 10^{5}~pb^{-1}$.

\begin{figure}[hbt]

\vspace*{0.5cm}
\hspace*{0.0cm}
\epsfxsize=10cm \epsfbox{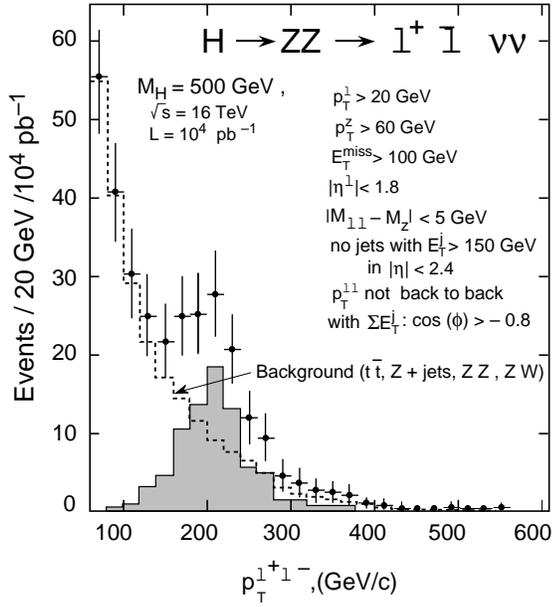}
\vspace*{0.0cm}

\caption[] {\label{fg:llnunu} \it $h \rightarrow l^+l^-\nu\nu$
signal  for $m_h~=~500~GeV$ with
$10^4pb^{-1}$ in CMS ($H \equiv h$).}
\end{figure}


\begin{figure}[hbt]

\vspace*{0.5cm}
\hspace*{0.0cm}
\epsfxsize=10cm \epsfbox{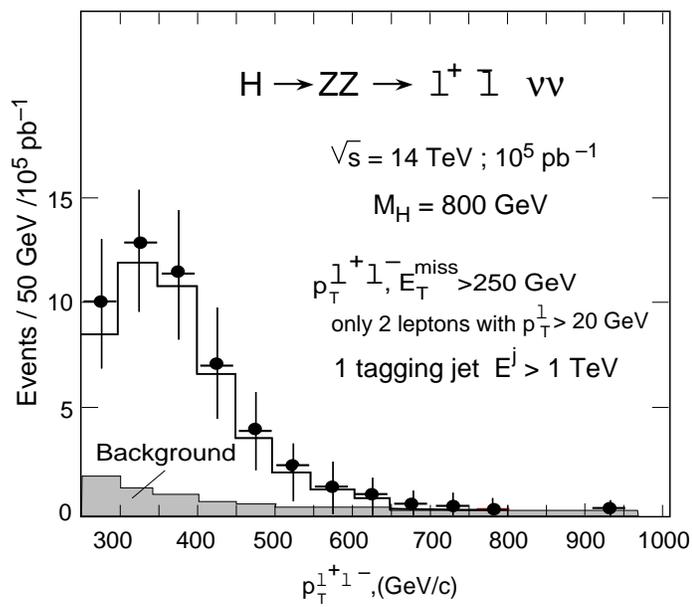}
\vspace*{0.0cm}

\caption[] {\label{fg:llnunu2} \it $h \rightarrow l^+l^-\nu\nu$
signal  for $m_h~=~800~GeV$ with
$10^5pb^{-1}$ in CMS. One tagged jet with $E>1~TeV$ is 
assumed ($H \equiv h$).}
\end{figure}


\subsection{Summary}

The most reliable signatures for the search for the Higgs boson at 
LHC are the following;

1. $h \rightarrow \gamma \gamma$ or $h \rightarrow \gamma\gamma + ~jets$

2. $h \rightarrow ZZ^{*},ZZ \rightarrow 4~l^{\pm}$

3. $h \rightarrow W^+W^- \rightarrow l^+ \nu l^- \bar{\nu}$

4. $h \rightarrow ZZ,WW \rightarrow ll \nu \nu,\, lljj,\,l \nu jj$

Fig.24 -25  show the expected CMS discovery potential of the standard 
Higgs boson as a function of $m_h$ for integrated luminosities of 
$10^5~pb^{-1}$ and $3\cdot 10^{4}~pb^{-1}$. For $L =10^{5}~pb^{-1}$ 
CMS is able to discover the Higgs boson at $ \geq 5\sigma$ level 
for the entire mass region $(95 ~GeV - 1~TeV)$. For low luminosity 
stage with $L = 3\cdot 10^{5}~pb^{-1}$ CMS is able to discover Higgs 
boson with a mass up to $\sim 600~GeV$.

\begin{figure}[hbt]

\vspace*{0.5cm}
\hspace*{0.0cm}
\epsfxsize=10cm \epsfbox{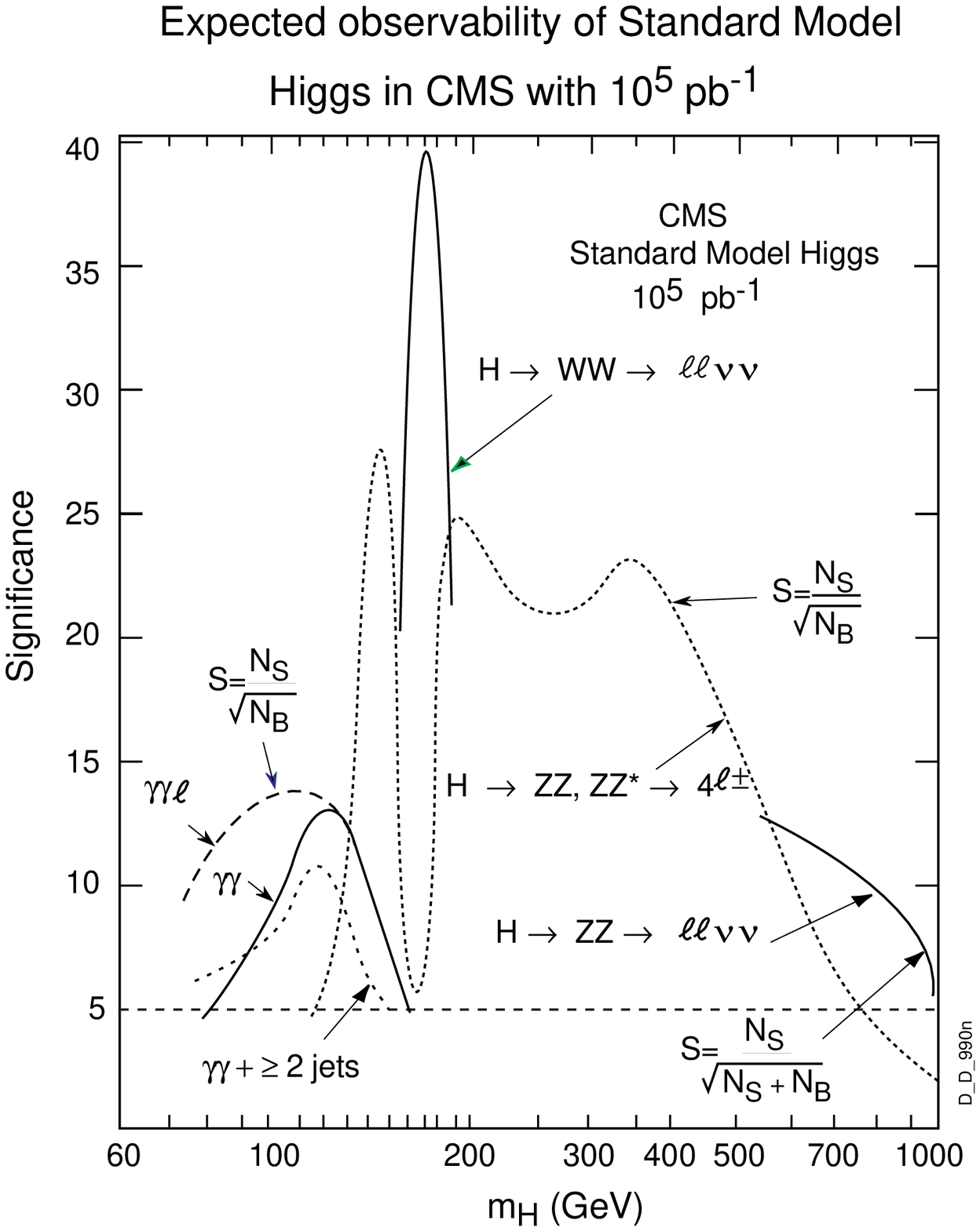}
\vspace*{0.0cm}

\caption[] {\label{fg:kun1} \it Expected observability of Standard Model 
Higgs as a function on $m_H$ in CMS with $10^5pb^{-1}$ (ref.\cite{79}) 
($H \equiv h$).}
\end{figure}


\begin{figure}[hbt]

\vspace*{0.5cm}
\hspace*{0.0cm}
\epsfxsize=10cm \epsfbox{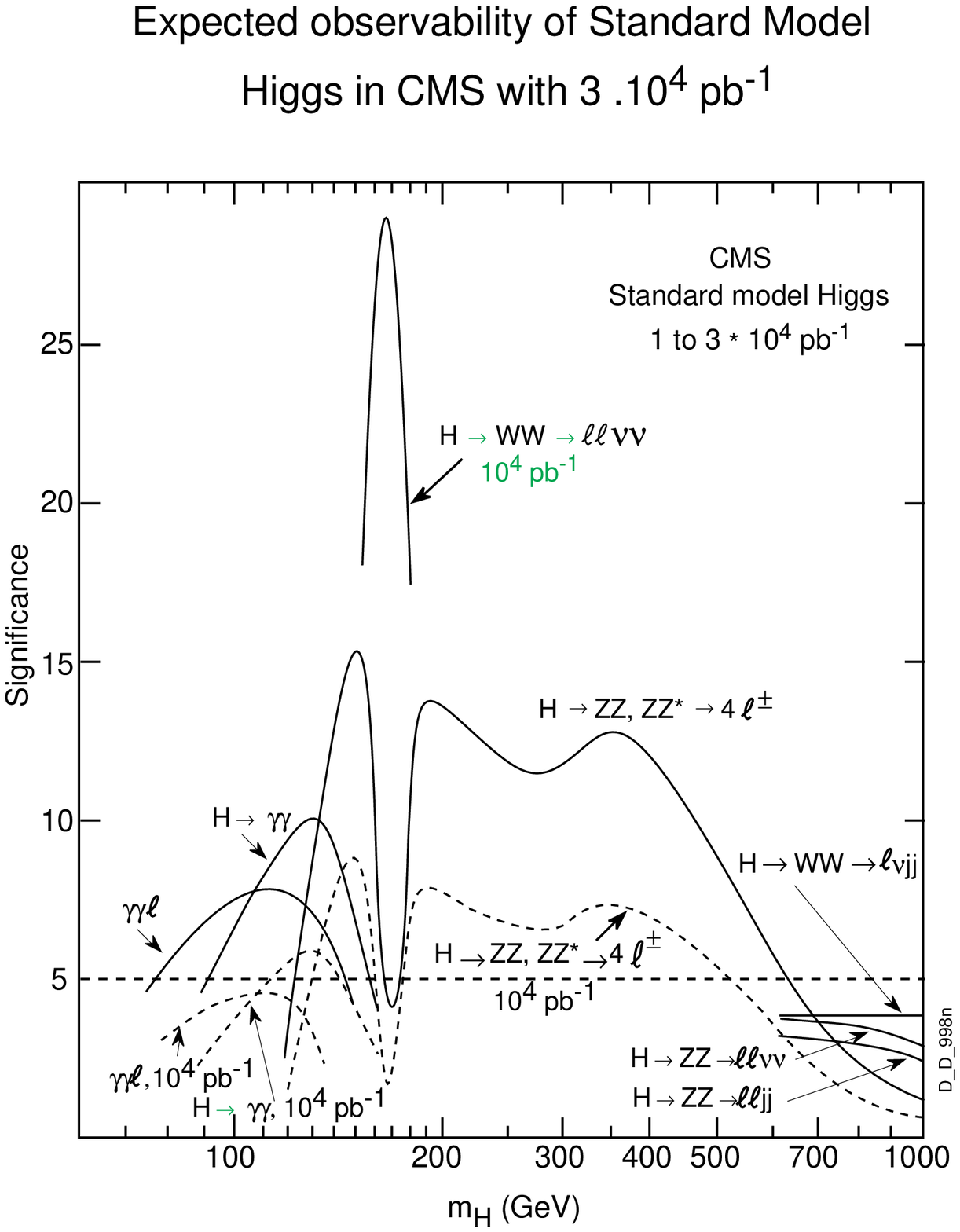}
\vspace*{0.0cm}

\caption[] {\label{fg:kun2} \it Expected observability of Standard Model 
Higgs as a function on $m_H$ in CMS with $3 \cdot 10^4pb^{-1}$ and
with $10^4pb^{-1}$ (ref.\cite{79}) ($H \equiv h$).}
\end{figure}


\newpage

\section{Conclusion}

There are no doubts that at present the supergoal number one 
of the experimental high energy physics is the 
search for the Higgs boson - the last non discovered 
cornerstone of the Standard Model. At present the LEP2 experimental
bound on the Higgs boson mass is $m_h \geq 102.6 ~GeV$. In a year 
LEP2 will be able to discover the Higgs 
boson or to increase a lower bound up to $(105-110) ~GeV$. LHC is able to 
discover the Higgs boson with a mass up to $1~TeV$ and to check 
its basic properties. The experimental Higgs boson discovery will be 
triumph of the idea of the renormalizability (in some sense it 
will be the ``experimental proof'' of the renormalizabilty of the 
electroweak interactions) which mathematical cornerstone is the 
famous Bogoliubov-Parasiuk theorem.  
At any rate after LHC we will know the basic mechanism 
(Higgs boson or something more exotic?)
of  the electroweak symmetry breaking.

We thank  our colleagues from INR theoretical department
for useful discussions. We are indebted to S.I.Bityukov for a help in 
preparation of the manuscript. The research described in this publication
has been supported by RFFI grant 99-02-16956.


\newpage
\begin{thebibliography}{99}

\bibitem{1} N.N.Bogolyubov and D.V.Shirkov, Introduction to the 
Theory of Quantized Fields \\
(3rd ed.) (John Wiley Inc., New York, 1980).
\bibitem{2} N.N.Bogolyubov and D.V.Parasyuk, Dokl.Akad.Nauk SSSR 
{\bf 100}(1956)429;  \\ 
Acta Mathem. {\bf 97}(1957)227.
\bibitem{3} N.N.Bogolyubov, A.A.Logunov, A.L.Oksak and I.T.Todorov, \\
General Principles of Quantum Field Theory, (Nauka, Moscow) 1987.
\bibitem{4} S.L.Glashow, Nucl.Phys.{\bf 22}(1961)579; \\  
S.Weinberg, Phys.Rev.Lett. {\bf 19}(1967)1264; \\
A.Salam, Elementary Particle Theory (ed. N.Svartholm) Almquist and 
Wiksells, Stockholm, 1964. 
\bibitem{5} P.Higgs, Phys.Lett. {\bf 12}(1964)132; F.Englert and R.Brout, \\
Phys.Rev.Lett. {\bf 13}(1964)321.
\bibitem{6} N.N.Bogolyubov, J.Phys. USSR {\bf 11}(1947)23; Lectures on \\ 
Quantum Statistics , 
Mackdonald Technical and Scientific, London, 1970.
\bibitem{7} E.Accomando et al., Phys.Rept. {\bf 299}(1998)1.
\bibitem{8} V.Barger,M.S.Berger, J.F.Gunion and T.Han,\\
Phys.Reports {\bf 286}(199701.  
\bibitem{9} Reviews and original references can be found in: \\
R.Barbieri, Riv.Nuovo Cim. {\bf 11}(1988)1; A.B.Lahanus and \\
D.V.Nanopoulos, Phys.Rep. {\bf 145}(1987)1; H.E.Haber and G.L.Lane, \\
Phys.Rep. {\bf 117}(1985)75; H.P.Nilles, Phys.Rep. {\bf 110}(1984)1.
\bibitem{10} L.B.Okun, Leptons and Quarks, ( North Holland 
Pub. Comp. 1982).
\bibitem{11} Ta-Pei Cheng and Ling-Fong-Li, Gauge \\
Theory of Elementary Particle Physics(Oxford University Press, \\
Oxford, 1984). 
\bibitem{12}  Stefan Pokorsky, Gauge Field Theories (Cambridge \\
University Press, Cambridge, 1987).
\bibitem{13} D.Bailin and A.Love, Introduction to Gauge Field Theory \\
(Adam Hilger, Bristol, 1986).
\bibitem{14} J.F.Gunion, H.E.Haber, G.Kane and \\
S.Dawson, The Higgs Hunter's Guide(Addison-Wesley Publishing \\
Company, Redwood City, CA)1990.
\bibitem{15}  V.I.Borodulin, R.N.Rogalyov and S.R.Slabospitsky,\\ 
 COmpendium of RElations, IHEP Preprint 95-50.
\bibitem{16} M.Spira and P.M.Zervas, Electroweak Symmetry Breaking and \\
Higgs Physics, CERN-TH/97-379 [hep-ph/9803257].
\bibitem{17} M.Spira, QCD Effects in Higgs Physics, CERN-TH/97-68  \\ 
(hep-ph/9705337), Fortsch.Phys. {\bf 43}(1998)203.
\bibitem{18} M.Dittmar, Searching for the Higgs and other Exotic Objects, \\
CMS CR 1999/009; ETHZ-IPP PR-98-10.
\bibitem{19} Review of Particle Physics, The European Physical Journal  \\ 
{\bf v.3}(1998)95.  
\bibitem{20} B.W.Lee, C.Quigg and C.B.Thacker, Phys.Rev.Lett. {\bf 38} \\
(1977)883; Phys.Rev. {\bf D10}(1974)1145.
\bibitem{21} S.Dawson and S.Willenbrock, Phys.Rev. {\bf D40}(1989)2880.
\bibitem{22} See for instance: \\ 
J.Jersak, in Higgs Particles(s), Proceedings of the 
Eighth INFN Eloisatron 
Project Workshop, \\ 
July 15-26, Erice, Italy, edited by A.Ali (Plenum Press, New York, 1990)p.39.
\bibitem{23} N.Cabibbo, L.Maiani, G.Parisi and R.Petronzio, Nucl.Phys. \\
{\bf B158}(1979)295; M.Lindner, Z.Phys. {\bf C31}(1986)295.
\bibitem{24} N.V.Krasnikov, Sov.J.Nucl.Phys. {\bf 28}(1978)549;P.Q.Hung, \\
Phys.Rev.Lett. {\bf 42}(1979)873; H.D.Politzer and S.Wolfram, \\
Phys.Lett. {\bf B82}(1979)242; A.A.Anselm, JETP Lett. {\bf 29} \\
(1979)590; M.Lindner, M.Sher and M.Zaglauer,Phys.Lett. \\
 {\bf B228}(1989)139.
\bibitem{25} N.V.Krasnikov, G.Kreyerhoff and R.Rodenberg, \\
Mod.Phys.Lett. {\bf A9}(1994)3663. 
\bibitem{26} J.Ellis, G.Ridolfi and F.Zwirner, Phys.Lett. {\bf B257} \\
(1991)83; H.Haber and R.Hempfling, Phys.Rev.Lett. {\bf 66}(1991)1815; \\
A.Yamada, Phys.Lett. {\bf B263}(1991)233; R.Barbieri, M.Frigeni and \\
F.Caravaglis, Phys.Lett. {\bf B258}(1991)233; P.M.Chanowski, \\
S.Pokorski and J.Rosick, Phys.Lett. {\bf B275}(1992)191.
\bibitem{27} N.V.Krasnikov and S.Pokorski, Phys.Lett. {\bf B288}(1992)184; \\
M.A.Diaz, T.A. ter Veldhuis and T.J.Weiler, Phys.Rev.Lett. \\ 
{\bf74}(1995)2876  and Phys.Rev. {\bf D54}(1996)5855.  
\bibitem{28} M.Veltman, Acta Phys.Polon. {\bf B8}(1977)475;
See also: \\ 
S.Dittmaier, D.Schildknecht and S.Weiglein, \\
Phys.Lett. {\bf B386}(1996)247.
\bibitem{29} A Combination of Preliminary Electroweak Measurements and 
Constraints \\
on the Standard Model, the LEP collaborations, LEPEWWG/98-01, 15 May 1998.
\bibitem{30} M.S.Chanowitz, Higgs boson mass constraints from precision \\
data and direct searches, LBNL-42103(1998)[hep-ph/9807452]  
\bibitem{31} L.Resnick, M.K.Sundarsean and P.J.S.Watson, Phys.Rev. {\bf D8}
(1973) 172; \\
J.Ellis, M.K.Gaillard and D.V.Nanopoulos, Nucl.Phys. {\bf B106}(1976)292.
\bibitem{32} E.Braaten and J.P.Leveille, Phys.Rev. {\bf D22}(1980)715; \\
N.Sakai, Phys.Rev. {\bf D22}(1980)2220; \\
M.Drees and K.Hikasa, Phys.Rev. {\bf D41}(1990)1547; \\ 
A.L.Kataev and V.T.Kim, Mod.Phys.Lett. {\bf A9}(1994)1309; \\
K.G.Chetyrkin, Phys.Lett. {\bf B390}(1997)309.
\bibitem{33} N.Cray, D.J.Broadhurst, W.Grafe and k.Schilcher,  \\
Z.Phys. {\bf C48}(1990)673.  
\bibitem{34} J.Fleischer and F.Jegerleher, Phys.Rev. {\bf D23}(1981)2001; \\
D.Yu Bardin, B.M.Vilenski and P.Kh.Khristova, Sov.J.Nucl.Phys. {\bf 53} 
(1991)152; \\
A.Dabelstein and W.Holik, Z.Phys. {\bf C53}(1992)507; \\
B.A.Kniehl, Nucl.Phys. {\bf B376}(1992)3. \\
\bibitem{35} B.A.Kniehl and M.Spira, Z.Phys. {\bf C69}(1995)77; \\
B.A.Kniehl and M.Spira, Nucl.Phys. {\bf B443}(1995)37.
\bibitem{36} A.Ghinculov, Nucl.Phys.{\bf B455}(1995)21; A.Frink, B.Kniehl, 
D.Kreimer, \\
and K.Riesselmann, Phys.Rev.{\bf D54}(1996)4548.
\bibitem{37} T.G.Rizzo, Phys.Rev. {\bf D22}(1980)389; W.-Y.Keung and 
W.J.Marciano,  \\
Phys.Rev. {\bf D30}(1984)248. 
\bibitem{38} R.N.Cahn, Rep.Prog.Phys. {\bf 52}(1989)389.   
\bibitem{39} J.Ellis, M.K.Gaillard and D.V.Nanopoulos, Nucl.Phys. {\bf B106}
(1976)292. 
\bibitem{40} M.Spira, A.Djouadi, D.Graudenz and P.M.Zervas, \\
Nucl.Phys.{\bf B453}(1995)17; \\
T.Inami, T.Kubotta and Y.Okada, Z.Phys. {\bf C18}(1983069; \\
A.Djouadi, M.Spira and P.M.Zervas, Phys.Lett. {\bf B264}(1991)440. 
\bibitem{41} K.G.Chetyrkin, B.A.Kniehl and M.Steinhauser, 
Phys.Rev.Lett. {\bf 79}(1997)353.
\bibitem{42} A.Djouadi and P.Gambino, Phys.Rev.Lett. {\bf D49}(1994)3499.   
\bibitem{43} H.Zheng and D.Wu, Phys.Rev. {\bf D42}(1990)3760; \\
A.Djouadi, M.Spira, J. van der Bij and P.M.Zervas, Phys.Lett. {\bf B257}
(1991)187; \\
S.Dawson and R.P.Kauffman, Phys.Rev. {\bf D47}(1993)1264.
\bibitem{44} J.Korner, K.Melnikov and O.Yakovlev, Phys.Rev. {\bf D53}
(1996)3737; \\
Y.Liao and X.Li, Phys.Lett. {\bf B396}(1997)225.
\bibitem{45} B.L.Ioffe and V.A.Khoze, Sov. J. Part. Nucl. {\bf 9}(1978)50; \\
J.D.Bjorken, Proc. Summer Institute on Particle Physics, 
Report SLAC-198(1976).
\bibitem{46} See for example: G.Altarelli, R.Kleiss and \\
C.Verzegnassi, editors, Physics at LEP, vol. 1: Standard Physics, \\
CERN Yellow Report 86-02(1986).
\bibitem{47} A.Blondel, Precision Electroweak Physics at LEP, \\
CERN-PPE/94-133(1994).
\bibitem{48} Report CERN 96-01, Vol. 1, ``Physics at LEP2', edited by 
G.Altarelli, \\
T.Sjostrand and F.Zwirner and references therein.
\bibitem{49} G.Passarino, Nucl.Phys. {\bf B488}(1997)3.
\bibitem{50} L3 Collaboration, Search for the Standard Model Higgs 
boson in $e^+e^-$ interactions at $\sqrt{s} =189~GeV$, CERN Preprint 
CERN-EP/99-080(1999).
\bibitem{51} The Opal Collaboration, Search for Neutral Higgs Bosons in 
$e^+e^-$ Collisions at $\sqrt{s} \approx 189~GeV$, CERN Preprint 
CERN-EP/99-09691999) [hep-ex/9908002]
\bibitem{52} The ALEPH Collaboration, Search for the Neutral Higgs 
bosons of the Standard Model and the MSSM in $e^+e^-$ Collisions at 
$\sqrt{s} =188.6~GeV$, ALEPH 99-053(CONF 99-029).
\bibitem{53} The DEPPHI Collaboration, DELPHI 99-8(CONF 208). 
\bibitem{54}M.Felcini, The Search for Higgs particles at LEP, 
hep-ex/9907049(1999). 
\bibitem{55} The ALEPH  Collaboration, Search for neutral Higgs bosons 
in $e^+e^-$ collisions at $\sqrt{s} \leq 196~GeV$, hep-ex/9908016(1999).
\bibitem{56} P. Mcnamara, Talk given at LEP EXPERIMENTS COMMITTEE; \\
CERN, 7 september 1999.
\bibitem{57} E.Gross et al., Prospects for the Higgs boson search in 
Electron-Positron collisions at LEP 200, CERN-EP/98-094.
\bibitem{58} H.Georgi S.L.Glashow, M.Machacek and D.V.Nanopoulos, \\ 
Phys.Rev.Lett. {\bf 40}(1978)692.
\bibitem{59} A.Djouadi, M.Spira and P.M.Zervas, Phys.Lett.{\bf B264} \\
(1991)440;  M.Spira, A.Djouadi, D.Graudenz and P.M.Zervas, \\
Nucl.Phys. {\bf B453}(1995)17. 
\bibitem{60} R.N.Cahn and S.Dawson, Phys.Lett. {\bf B136}(1984)196; \\
K.Hikasa, Phys.Lett.{\bf B164}(1985)341; \\
G.Altarelli, B.Mele and F.Pitolli, Nucl.Phys. {\bf B287}(1987)205; \\
T.Han, G.Valensia and S.Willenbrock, Phys.Rev.Lett. {\bf 69}(1992)3274.
\bibitem{61} S.L.Glashow, D.V.Nanopoulos and A.Yildiz, 
Phys.Rev. {\bf D18}(1978)1724.
\bibitem{62} Z.Kunszt, Nucl.Phys. {\bf B247}(1984)339; \\
J.F.Gunion, Phys.lett. {\bf B253}(1991)269; \\
W.J.Marciano and F.E.Paige, Phys.Rev.Lett. {\bf 66}(1991)2433.
\bibitem{63} C.Quigg, Physics Opportunities in Fermilab's Futures, \\ 
FERMILAB-FN-676, March 1999. 
\bibitem{64} Hugh E. Montgometry, Physics with the Main injector, \\
hep-ex/9904019. 
\bibitem{65} Tao Han, Andre S.Turcot and Ren-Jie Zhang, Exploiting  \\
 $h \rightarrow W^{*}W^{*}$ Decays at the Upgraded Fermilab Tevatron, \\ 
 FERMILAB, MADPH-08-1094(1998), hep-ph/9812275(1998).  
\bibitem{66} The Large Hadron Collider, CERN/AC/95-05. 
\bibitem{67} CMS, Technical Proposal, CERN/LHCC/94-38 LHCCP1, 15 december 1994.
\bibitem{68} ATLAS, Technical Proposal, CERN/LHCC/94-43 LHCCP2, 
15 december 1994
\bibitem{69} As a review of physics to be studied at LHC see, for example: \\
N.V.Krasnikov and V.A.Matveev, Phys.Part.Nucl. {\bf 28}(1997)441.  
\bibitem{70} C.Seez, $H \rightarrow \gamma \gamma$; An update, 
CMS TN/94-289(1994).
\bibitem{71} K.Lassila-Perrini, The reconstruction of $ Higgs 
\rightarrow \gamma \gamma$ in CMS, CMS CR/97-006(1997).
\bibitem{72} S.Abdullin, A.Starodumov and N.Stepanov, Study of the Associated 
Production Modes $WH$ and $t\bar{t}H$ in CMS, CMS TN/93-86(1993).
\bibitem{73} S.Abdullin, A.Knezevic, R.Kinnunen and N.Stepanov, Possibilities 
to improve the observability of SM light Higgs in the $\gamma\gamma$ 
channel, CMS TN/94-247(1994).
\bibitem{74} I.Iashvili, R.Kinnunen, A.Nikitenko and D.Denegri, 
Study of the $H \rightarrow ZZ^{*} \rightarrow 4l^{\pm}$ Channel in CMS, 
CMS TN/95-059(1995).
\bibitem{75}  C.Charlot, A.Nikitenko, I.Puljak and I.Soric, Comparison of 
fixed window \\ 
and clusterizarion algorithms for $Z \rightarrow e^+e^-$ and 
$H \rightarrow 4e^{\pm}$ in CMS \\
$PbWO_4$ crystal ECAL for Higgs mass 170 and 130 $GeV$, CMS TN/95-101(1995).
\bibitem{76} D.Bomestar, D.Denegri, R.Kinnunen and A.Nikitenko, 
Study of the $H \rightarrow ZZ \rightarrow 4l^{\pm}$ \\ 
with full GEANT simulation of CMS detector, CMS TN/94-018(1994).
\bibitem{77} M.Dzelalija, Z.Antonovic and R.Kinnunen, Study of the 
heavy \\ 
$H \rightarrow ZZ \rightarrow 4l^{\pm} $ in CMS, CMS TN/95-076(1995).
\bibitem{78} S.Abdullin and N.Stepanov, Towards self-consistent scenario of 
the Heavy Higgs \\ 
observabilty via the channels $l\nu jj$ and $lljj$ at 
CMS, CMS TM/94-178(1994). 
\bibitem{79} R.Kinnunen and D.Denegri, Expected SM/SUSY Higgs observability 
in CMS, CMS Note 1997/057(1997). 
\bibitem{80} V.Drollinger, T.Muller and R.Kinnunen, Possibilities 
of $t\bar{t} H^0$ Event Reconstruction, CMS Note 1999/001(1999).
\bibitem{81} M.Spira and M.Dittmar, Standard Model Higgs cross 
sections(NLO) and PYTHIA, CMS Note 1997/080(1997).
\bibitem{82} M.N.Dubinin, V.A.Ilyin and V.I.Savrin, Light Higgs Boson 
signal at LHC in the Reactions \\
$pp \rightarrow \gamma\gamma +jet$ and 
$pp \rightarrow \gamma\gamma + lepton$, CMS Note 1997/101; \\
S.Abdullin, M.Dubinin, V.Ilyin, D.Kovalenko, V.Savrin \\
and N.Stepanov, Phys.Lett.{\bf B431}(1998)410. 
\bibitem{83} M.Dittmar and H.Dreiner, Phys.Rev. {\bf D55}(1997)167
[hep-ph/9608317]; \\
LHC Higgs Search with $l^+{\nu} l^-\bar{\nu}$ final states, CMS Note 1997/083.
\bibitem{84} D.Green et al., Search for the Standard Model Higgs Boson with 
$M_H \approx 170~GeV/c^2$ in $W^+W^-$ Decay Mode, CMS Note 1998/089. 
\bibitem{85} I.Iashvili, R.Kinnunen, A.Nikitenko and D.Denegri, Study of the 
$H \rightarrow 4l^{\pm}$ Channel in CMS, CMS Note 1995/059. 
\bibitem{86} N.Stepanov, Search for Heavy Higgs via the 
$H \rightarrow ll\nu\nu $ Channel, CMS TN/93-87(1993); \\
N.Stepanov and A.Starodumov, Search for Higgs in the TEV Region, 
CMS-TN/92-49(1992). 
\bibitem{87} S.Zmushko, D.Frodeavaux and L.Poggioli, $H \rightarrow WW 
\rightarrow l\nu jj$ and $H \rightarrow ZZ \rightarrow lljj$. Particle 
level studies, ATLAS Internal Note PHYS-No-103(1997).
\bibitem{88} L.Poggioli, $H \rightarrow ZZ^{*} \rightarrow 4~leptons$. 
A comparison of ATLAS and CMS potentials, ATLAS Internal Note, 
PHYS-No-066(1995). 
\bibitem{89} D.Froidevaux, F.Gianotti and E.Richter-Was, Comparison \\     
of the ATLAS and CMS discovery potential for the 
$H \rightarrow \gamma\gamma$ \\  
channel at the LHC, ATLAS Internal Note PHYS-No-64(1995).
\bibitem{90} V.Cavasinni, D.Costanzo, S.Lami and F.Spano, Search for 
$H \rightarrow WW \rightarrow l\nu jj$ with the ATLAS Detector ($m_H = 
300-600~GeV$, ATLAS Internal Note,ATL-PHYS-98-127(1998).
\bibitem{91} O.Linossier and L.Poggioli, $H^0 \rightarrow ZZ^{*} 
\rightarrow 4l$ channel, in ATLAS. Signal reconstruction and 
Reducible backgrounds rejection, ATLAS Note Phys No 101(1997).
\bibitem{92} S.Zmushko, $H \rightarrow \gamma\gamma$ 
in association with jets, ATLAS Internal Note ATL-PHYS-99-009(1999).
\bibitem{93} E.Richter-Was and M.Sapinski, Search for the SM and MSSM Higgs \\
boson in the $t\bar{t}H, \, HH \rightarrow b\bar{b}$ channel, \\ 
ATL-PHYS-98-132(1998).
\bibitem{94} P.Savard and G.Azuelos, The discovery potential \\ 
of a Heavy Higgs ($M_H = 800~GeV$) using full GEANT sumulations \\ 
of ATLAS, ATL-PHYS-98-128(1998).
\bibitem{95} O.Linossier and R.Zitoun, $H_0 \rightarrow ZZ^{*} 
\rightarrow 4~l \, channel$, \\ 
in ATLAS - A complementary study of the $ZZ^{*}$ background, \\ 
ATL-Phys-96-096(1996).
\bibitem{96} V.Tisserand, The Higgs to two photon decay  \\ 
in the ATLAS detector UPDATED, ATL-PHYS-96-091.
\bibitem{97} S.I.Bityukov and N.V.Krasnikov, Mod.Phys.Lett. 
{\bf A13}(1998)3235;  \\ 
S.I.Bityukov and N.V.Krasnikov, Observability and Probability of Discovery \\ 
in Future Experiments, hep-ph/9908402(1999), CMS IN Note 1999/027.  
\end{thebibliography}
\end{document}